\pdfoutput=1

\documentclass{aastex631}
\usepackage{epstopdf}
\usepackage{xspace}
\usepackage{graphicx}
\usepackage{amssymb}
\usepackage{amsmath}
\usepackage{float}
\usepackage{hyperref}

\shorttitle{Estimating the Ambient Medium Density Around Radio Galaxies}
\shortauthors{W\'ojtowicz et al.}

\begin{document}

\title{A Novel Method for Estimating the Ambient Medium Density Around Distant Radio Sources\\ from Their Observed Radio Spectra}

\correspondingauthor{Anna W\'ojtowicz}
\email{awojtowicz@oa.uj.edu.pl}

\author{Anna W\'ojtowicz}
\affiliation{Astronomical Observatory of the Jagiellonian University, ul. Orla 171, 30-244 Krak\'ow, Poland}

\author{{\L}ukasz Stawarz}
\affiliation{Astronomical Observatory of the Jagiellonian University, ul. Orla 171, 30-244 Krak\'ow, Poland}

\author{Jerzy Machalski}
\affiliation{Astronomical Observatory of the Jagiellonian University, ul. Orla 171, 30-244 Krak\'ow, Poland}

\author{Luisa Ostorero}
\affiliation{Dipartimento di Fisica -- Universit\`a degli Studi di Torino and Istituto Nazionale di Fisica Nucleare (INFN), Via P. Giuria 1, I-10125 Torino, Italy}
\affiliation{Istituto Nazionale di Fisica Nucleare (INFN), Sezione di Torino,  Via P. Giuria 1, I-10125 Torino, Italy}

\begin{abstract}
The dynamical evolution and radiative properties of luminous radio galaxies and quasars of the FR\,II type, are well understood. As a result, through the use of detailed modeling of the observed radio emission of such sources, one can estimate various physical parameters of the systems, including the density of the ambient medium into which the radio structure evolves. This, however, requires rather comprehensive observational information, i.e. sampling the broad-band radio continua of the targets at several frequencies, and imaging their radio structures with high resolution. Such observations are, on the other hand, not always available, especially for high-redshift objects. Here we analyze the best-fit values of the source physical parameters, derived from an extensive modeling of the largest currently available sample of FR\,II radio sources, for which good-quality multi-wavelength radio flux measurements could be collected. In the analyzed dataset, we notice a significant and non-obvious correlation between the spectral index of the non-thermal radio emission continuum, and density of the ambient medium. We derive the corresponding correlation parameters, and quantify the intrinsic scatter by means of Bayesian analysis. We propose that the discovered correlation could be used as a cosmological tool to estimate the density of ambient medium for large samples of distant radio galaxies. Our method does not require any detailed modeling of individual sources, and relies on limited observational information, namely the slope of the radio continuum between the rest-frame frequencies 0.4\,GHz and 5\,GHz, possibly combined with the total linear size of the radio structure.
\end{abstract}
    

\section{Introduction}
\label{sec:intro}

High-power radio galaxies are believed to constitute a parent population of radio-loud quasars \citep[e.g.,][]{Barthel89}. Their large-scale radio structures are typically characterized by the edge-brightened morphology of ``classical-double'' type, consisting of a pair of well-collimated jets terminating in bright, distinct hotspots, and surrounded by extended lobes that dominate the radiative outputs of systems at radio wavelengths \citep[hereafter ``FR\,II'' sources, following][]{fan74}. There is an ongoing debate on whether this large-scale morphology is determined solely by the kinetic power of the jets (and so, whether it depends only on the total radio power of a source), or also by the type of host galaxy \citep[e.g.,][]{led96,koz11,win11,cap17,min19}.

Furthermore, high-power radio galaxies at lower redshifts seem to avoid dense large-scale environments, especially when compared with local low-power radio galaxies that are often found in the centers of rich clusters \citep[e.g.,][and references therein]{zir97,har02,win11}. For example, a careful examination of the NRAO\footnote{The National Radio Astronomy Observatory is a facility of the National Science Foundation operated under cooperative agreement by Associated Universities, Inc.} Very Large Array (VLA) imaging survey of Abell clusters at northern declinations \citep{owe92,owe97}, when restricted to the systems with richness $R \geq 0$ and redshifts $z < 0.25$, returns only two FR\,IIs hosted by the ``brightest cluster galaxies'' (\citealt{Stawarz14,Hagino15}; see also \citealt{Cheung19}). In general --- i.e., regardless of the classification of the host as BCG --- there still is a significant difference in the environment of galaxies displaying different radio morphologies. For example, very recently \citet{Croston19} have demonstrated that, among radio galaxies at $z <0.4$, FR\,IIs are characterized by a lower cluster association fraction, and systematically poorer environments.

Whether the difference in the large-scale environment of low- and high-power radio galaxies in the local Universe is significant or not, and whether this difference persists also in the higher-$z$ Universe, both remain open questions \citep[see in this context, e.g.,][]{hl91,bel07,ant12,mas19,mas20}. Part of the problem here is that, the X-ray observations of a sufficient depth and quality to allow one to estimate --- via the detection of a thermal (free-free and line) emission component --- the temperature and spatial distribution of the hot gaseous fraction of the intergalactic medium (IGM), are currently still rather sparse, and restricted to the nearest and/or the brightest systems \citep[see][]{Belsole07,Ineson17}.

In this paper, we explore the possibility of using high-power FR\,II radio galaxies as cosmological probes of their environment, and in particular, of using basic observed properties of their radio emission continua to estimate the gas density of the ambient medium into which their radio structures evolve.\footnote{We note that, very recently, \citet{Turner20} proposed a method which, through Bayesian inference, provides probability density functions for the most likely redshift of radio galaxies, based on the radio-frequency imaging and photometry.} Note in this context, that since FR\,IIs are expected to reside in the external regions of galaxy clusters, as mentioned in previous paragraphs, an estimate of the surrounding gas density would likely yield an estimate of the density of a galaxy group that is falling into the cluster. The external regions of galaxy clusters are indeed rather in-homogeneous, in terms of both galaxy density and gas density \citep[e.g.,][]{Rines02,Coe12,Geller14,Yu16,Haines18,Sohn19}. A typical elliptical host of an FR\,II source \citep[e.g.,][and references therein]{Sikora07,Zheng20}, is likely to be the brightest galaxy member of such a small group \citep[see][]{Lin10} that is captured by a cluster.

In principle, the general idea that FR\,IIs could be used in this or another way as probes of the distant Universe, may not sound surprising or novel. Indeed, the dynamical evolution of such sources is rather well-understood \citep[see][and the next section]{beg89,KA,kaw06}, as is the production of non-thermal emission in their expanding radio lobes \citep[e.g.,][]{KDA,stawarz08}. In the framework of this relatively simple scenario, the total kinetic power of the jets, along with properties of the ambient medium (density and pressure profiles), are key factors, determining the volume, expansion velocities, and total energy content of the lobes, and hence the spectral evolution of the lobes' non-thermal emission. As a result, through detailed modeling of the observed radio emission continua of FR\,IIs, using approximate prescriptions that combine dynamical evolution of large-scale structures with the particle and radiative transfer equations, one could hope to constrain the main physical parameters of the model, such as the ambient gas density. This is, however, a rather time-consuming procedure, requiring, in addition, comprehensive observational information, and as such could hardly be used when dealing with large samples of sources at large cosmological distances emerging from new-generation massive radio surveys, such as the Low-Frequency Array (LOFAR) Two-Metre Sky Survey (LoTSS) survey \citep{Hardcastle19}.

Here we propose an alternative approach, which enables us to avoid detailed modeling of individual sources, and is based on limited observational information. This method instead exploits a non-obvious correlation emerging between the best-fit values of radio spectral index, determined between two rest-frame frequencies, and the density of the ambient medium --- hereafter referred to as \emph{the model data} --- derived from detailed dynamical-radiative modeling of a large sample of FR\,IIs with available multi-wavelength radio flux-density measurements. 

The paper is organized as follows: In Section\,\ref{sec:method}, we describe the dynamical-radiative model used to derive the model data; in Section\,\ref{sec:stat} we perform a correlation analysis on the model data, and the regression analysis by means of Bayesian approach; in Section\,\ref{sec:discussion}, we verify our findings by examining good-quality high-angular resolution X-ray data available for some of the sources from our sample; final conclusions are given in Section\,\ref{sec:conclusions}.

\section{Dynamical modeling}
\label{sec:method}

In this paper, we explore the parameter space emerging from modeling the largest sample of FR\,II radio sources selected from various radio catalogs and surveys, namely the Cambridge Catalogs of Radio Sources (3C--6C--5C), the Bologna Sky Surveys (B2--B3), and the Green Bank Sky Surveys (GB--GB2), as described extensively in Machalski et al. (2021, ApJS, in press). Keeping in mind that our dynamical-radiative modelling (see below) deals exclusively with the extended lobes of classical doubles, and requires rather good-quality radio photometry and imaging in a wider frequency range, the following criteria have been applied when selecting the targets: (i) We included only the lobe-dominated sources previously identified as FR\,IIs, i.e. those classical doubles with the core/jet emission accounting to $<5\%$ of the total emission at radio frequencies; in other words, all the core-dominated sources present in the considered radio catalogs and surveys, even if classified as FR\,IIs, have been excluded from the analysis. (ii) We included only the sources which are resolved on the available radio maps sufficiently well, so that the projected linear sizes and axial ratios of the radio lobes (assuming cylindrical geometries), could be measured. (iii) We considered only the sources with a robust optical identification of the host galaxy, with measured spectroscopic redshift. (iv) We required well characterized radio spectra of the sources, with flux densities measured at least at five observing frequencies, covering the frequency range between 74\,MHz and 5\,GHz, so that the two-point spectral index between the rest-frame frequencies 0.4\,GHz and 5\,GHz could be determined.

There were no limits assigned for the redshift range, radio luminosities, or linear sizes of the sources included in the final sample. The redshifts of the selected objects span the range from 0.03 up to 4.41, with a mean value of 0.77.

The preliminary sample analyzed here consists of 271 targets, including 188 sources selected from the 3CRR, 6CE, and 7CRS in the compilation of \citet{Grimes04}, and 69 low-luminosity B2 and B3 sources from \citet{deRuiter86} and \citet{Fanti86,Fanti87}, supplemented with GB/GB2 middle-luminosity sources from \citet{Machalski98}. Most of these sources ($>80\%$) are identified as radio galaxies, with only a minority ($<20\%$) classified spectroscopically as quasars.

For the modeling procedure, we use the DYNAGE algorithm of \citet{Machalski07}. This code has already been applied and verified several times to individual double radio sources belonging to various classes and types, such as giant radio galaxies, double-double radio galaxies, X-shaped radio galaxies, and luminous radio quasars \citep[e.g.,][]{Machalski10,Machalski11,Machalski16,Bhatta18,OSullivan19}. It is based on the \citet{KA} analytical model for the dynamical evolution of classical doubles, augmented by the prescription for energy evolution of the lobes' non-thermal emission by \citet{KDA}.

In particular, the \citet{KA} model describes a self-similar expansion of a radio cocoon powered over the time $\tau$ by a pair of jets with the total kinetic luminosity $Q$, and expanding into the hot gaseous environment whose density is modeled with a $\beta$ profile
\begin{equation}
\rho(r) = \rho_0 \times \left(r/a_0\right)^{-\beta} \, ,
\label{eq:rho}
\end{equation}
where $r > a_0$ is the radial distance from the center of the host galaxy. The entire bulk kinetic energy carried away by the jets from the radio core is converted, at the termination shocks (``hotspots''), into the internal energy of a magnetized plasma which, subsequently, inflates the radio cocoon, forming, in this way, extended radio lobes, and providing the pressure support for the cocoon's sideways expansion. At the shock front, the jets' particles attain a non-thermal energy distribution, approximated in the model by a single power-law with the energy index $s$, so that the corresponding ``injection'' spectral index of the hotspots' synchrotron emission (in the optically-thin segment of the spectrum) is $\alpha_{\rm inj} = (s-1)/2$. Further downstream the shock, i.e. deep within the cocoon, particles loose their energies due to radiative and adiabatic cooling; as a result, the radio continuum, representing the integrated radiative output of the entire lobes, steepens at higher frequencies to $\alpha > \alpha_{\rm inj}$. 

\begin{deluxetable}{cccccccc}[!t]
\tabletypesize{\scriptsize}
\tablecaption{Results of the correlation analysis and median fit parameters following from the Bayesian regression analysis \label{tab:results}}
\tablewidth{0pt}
\tablehead{\\
& \multicolumn{2}{c}{Correlation analysis} & & \multicolumn{4}{c}{Bayesian regression analysis} \\
\cline{2-3} \cline{5-8}
\colhead{Dataset} & \colhead{Pearson} & \colhead{Kendall} & \colhead{Model} & \colhead{$\tilde{a}$} & \colhead{$\tilde{b}$} & \colhead{$\tilde{c}$} & \colhead{$\tilde{\sigma}$}\\
\colhead{(1)} & \colhead{(2)} & \colhead{(3)} & \colhead{(4)} & \colhead{(5)} & \colhead{(6)} & \colhead{(7)} & \colhead{(8)}
}
\startdata
\\
$(\log \rho_i, \, \log D_i, \, \alpha_i)$ & $\rho=-0.785$ & $\tau=-0.593$ &    $\log\!\rho \sim$ & $0.28 \pm 0.23$ &  $-1.01 \pm 0.07$ & $3.25 \pm 0.22$  & $0.40\pm0.02$\\
 & $p<2.2e-16$ & $p<2.2e-16$ & $\mathcal{N}\!\left(a + b \, \log\!D + c \, \alpha,\sigma\right)$ & & & &\\
 \\
$(\log \rho_i, \, \alpha_i)$ & $\rho= 0.459$ & $\tau=0.328$  & $\log\!\rho \sim$ & $-1.51 \pm 0.29$ &  $2.65 \pm 0.33$ & & $0.58\pm0.03$\\
 & $p=1.4e-14 $ & $p=7.62e-15$ & $\mathcal{N}\!\left(a + b \, \alpha;\,\sigma\right)$ & & & &\\
 \\
$(\log\![\rho_0 a_0^{3/2}]_i, \, \log D_i, \, \alpha_i)$ & $\rho=0.757$ & $\tau=0.520$ &    $\log \rho_0 a_0^{3/2} \sim$ & $-3.97 \pm 0.22$ &  $0.50\pm 0.06$ & $3.26\pm0.23$  & $0.39\pm 0.02$\\
 & $p< 2.2e-16$ & $p< 2.2e-16$ & $\mathcal{N}\!\left(a + b \, \log\!D + c \, \alpha,\sigma\right)$ & & & &\\
 \\
$(\log\![\rho_0 a_0^{3/2}]_i, \, \alpha_i)$  &  $\rho=0.675$ & $\tau=0.452$ &    $\log \rho_0 a_0^{3/2} \sim$ & $-3.09^{+0.24}_{-0.23}$ &  $3.55^{+0.26}_{-0.27}$ & & $0.44\pm 0.02$\\
 & $p<2.2e-16$ & $p < 2.2e-16$ & $ \mathcal{N}\!\left(a + b \, \alpha;\,\sigma\right)$ & & & &\\
 \\
 \hline
 \enddata
\tablecomments{Col.\,(1): dataset; Col.\,(2): Pearson's product-moment correlation coefficient $\rho$ and the corresponding $p$-value; Col.\,(3): Kendall's rank correlation coefficient $\tau$ and the corresponding $p$-value; Col.\,(4): statistical model for the Bayesian regression analysis; Col.\,(5--7): coefficients of the linear relation; Col.\,(8): intrinsic spread related to the unspecified variables.}
\end{deluxetable}

The numerical code DYNAGE has been developed to solve the inverse problem, i.e., to determine the four main free parameters of the model --- namely, the jet kinetic power $Q$, the jet lifetime $\tau$, the injection index $\alpha_{\rm inj}$, and the central density of the gaseous environment $\rho_0$ --- by fitting the model to a given set of observables, including the linear size of the lobes, their volume, as well as the slope and normalization of the observed radio continuum. This fitting procedure yields a unique set of best-fit parameters. Obviously, the model adopts several crude approximations. For example, we assume the jet inclinations as either $70$\,deg or $90$\,deg for most of the radio galaxies in the sample, while in the case of quasars we accommodate even smaller values, all as guided by the lobe/counter-lobe radio appearance in individual targets (surface brightness asymmetry, etc.). Also, we consider a universal slope for the ambient gas density profile $\beta=3/2$. We emphasize, in this context, that the $\beta$-model for the IGM density distribution, assumes in general a hydrostatic equilibrium for the isothermal hot gas and galaxies \citep{King62,Cavaliere76}; when applied to galaxy clusters' X-ray surface brightness profiles, the resulting best-fit value for the $\beta$ parameter (as defined in equation\,\ref{eq:rho} above) are, indeed, typically within the range between 1.2 and 2.5 \citep[see, e.g., the recent analysis by][]{Kafer19}. We also note that, in the framework of the \citet{KA} model, in which the radio structure is assumed to evolve in a power-law external density profile, $a_0$ and $\rho_0$ are not independent parameters, and therefore the model fit returns in fact the best-fit value for the characteristic quantity $\rho_0 a_0^{\beta}$ rather than $\rho_0$ alone.

Most importantly, the model treats $\alpha_{\rm inj}$ as a \emph{free} parameter. This parameter should, in fact, be considered as an \emph{``effective''} injection spectral index, i.e. the injection index averaged over a broader spectral range and over the lifetime of a source, since the particle energy spectrum formed at the termination shocks may evolve with time, and may be far more complex than a single power-law \citep[see the discussion in][]{Machalski07}. The integrated spectral index of the lobes' radio emission, $\alpha$, within a given frequency range, depends obviously on $\alpha_{\rm inj}$, but also on the other model parameters that determine the energy evolution of the radiating particles within the lobes, $\alpha = f\!\left(\alpha_{\rm inj}, Q, \tau, \rho_0\right)$. In other words, the spectral curvature of the lobes' radio continuum at a given moment $\tau$, encodes the information on the injection index, as well as on the general evolutionary history of a source.

Within this framework note that, in the \citet{KA} model, the linear size of the source, $D$, scales with the source lifetime as $D \propto (Q/\rho_0 a_0^{\beta})^{1/(5-\beta)} \, \tau^{3/(5-\beta)}$, so that with the adopted $\beta=3/2$ the ambient medium density at distances $r \simeq D/2$
\begin{equation}
\log\!\rho \propto -5\,\log\!D+ \log\!Q+ 3\,\log\!\tau \, .
\label{eq:plane}
\end{equation}
On the right-hand side of the above relation, only the source linear size $D$ may be estimated directly from observations, while the jet power $Q$ and the jet lifetime $\tau$ can only be determined through a careful modeling by using algorithms such as DYNAGE. Yet, as mentioned in the previous paragraph, those unknown parameters determine, at the same time, the observable curvature of the ageing radio continuum of the lobes. It is thus reasonable to assume that the $\rho = f\!(D, Q, \tau)$ scaling is equivalent to the relation $\rho = f\!(D, \alpha, \alpha_{\rm inj})$. If this is indeed the case, the modeling of a large sample of sources would in principle enable to ``calibrate'' such a relation, obtaining at the end a simple correlation of the parameter we seek, $\log\!\rho$, solely with the observable $\log\!D$ and $\alpha$. The additional dependence on $\alpha_{\rm inj}$ would then manifest as a correlation spread (a spread which, on the other hand, should not be substantial as long as $\alpha$ and $\alpha_{\rm inj}$ are tightly correlated; see Appendix\,\ref{aa} for the additional details).

In the following Section, we demonstrate that in the parameter space emerging from the DYNAGE modeling of the compiled sample of FR\,II radio galaxies, not only does a well-defined correlation plane $(\log\!\rho, \log\!D, \alpha)$ exist, but in fact there is a much simpler correlation $\log\!\rho \propto \alpha$, albeit with a lower significance. By means of a Bayesian approach, we quantify the intrinsic spread of those correlations, and show that this spread decreases when $\log\!\rho_0$ (or rather $\log\!\rho_0 a_0^{\beta}$) is used in the regression analysis instead of $\log\!\rho$. 

The full sample of 271 targets, along with the corresponding model data, is presented in Appendix\,\ref{ab}. The statistical analysis given in the following Section\,\ref{sec:stat}, is however performed on a slightly smaller sample of 253 sources, formed after excluding all the objects with linear sizes $D> 1$\,Mpc. Indeed, we argue that for such giant radio sources, our simplified evolutionary scenario of a self-similar lobes' expansion in a $\beta=3/2$ ambient density atmosphere, may not be really justified anymore, since at such large $\mathcal{O}(1\,{\rm Mpc})$ distances from the galaxy group centers, hot gas may be far from the hydrostatic equilibrium condition \citep[e.g.,][and references therein]{Walker19}. In this context see also Appendix\,\ref{ac} and the discussion therein.

We finally note that, although the spectral index between the two chosen rest-frame frequencies could be determined directly from the observations of the selected targets, in our model data we use instead the values derived from the modeled spectra between the emitted frequencies 0.4\,GHz and 5\,GHz. 

\section{Statistical Analysis}
\label{sec:stat}

\subsection{Correlation Plane}
\label{sec:plane}

Let us assume a linear relation $y=y(\mathbf{X})$ between the ``response'' variable $y\equiv \log\!\rho$, and the ``prediction'' variables $\mathbf{X} \equiv (\log\! D, \alpha)$, 
\begin{equation}
    \log\!\rho = a + b \, \log\!D + c\,\alpha \,+ \epsilon ,
    \label{eq:relation}
\end{equation}
where $\boldsymbol{\theta} \equiv (a,b,c)$ is the set of the coefficients of the linear relation and $\epsilon\sim \mathcal{N}(0,\sigma^2)$ is the error term with a normal distribution and variance $\sigma$. Our training dataset $\mathcal{D}$ is the DYNAGE matrix $(\log \rho_i, \, \log D_i, \, \alpha_i)$, where $i=1,2,...,N$ with $N=253$, the source linear size, $D$, is expressed in kpc (so that $D<10^3$), and the ambient medium density, $\rho$, is given in the units of $10^{-28}$\,g\,cm$^{-3}$ (so that, for example, the value $\log\!\rho \simeq 0$ corresponds roughly to the gas number density $n \simeq \rho/m_p \sim 10^{-4}$\,cm$^{-3}$). 

The detaset displays a strong, statistically significant correlation, for which the $p$-values are $< 2.2 \times 10^{-16}$ for both Pearson's product-moment correlation test (returning the coefficient $\rho=-0.785$) and Kendall's rank correlation (returning $\tau=-0.593$), as summarized in Table\,\ref{tab:results}. However, as the relation expressed in equation\,\ref{eq:relation} follows from a much more complex model, that was briefly described in Section\,\ref{sec:method}, there may be embedded dependence on additional variables. We thus decided to find the linear relation coefficients $\boldsymbol{\theta}$ using the Bayesian approach to constrain the credible \emph{distributions} of the model correlation parameters. This approach enables us to quantify the intrinsic scatter arising  from the dependence on variables not taken into account in the correlation analysis.

\begin{figure}[!t]
\centering
\vspace{1.5cm}
\includegraphics[width=0.8\textwidth]{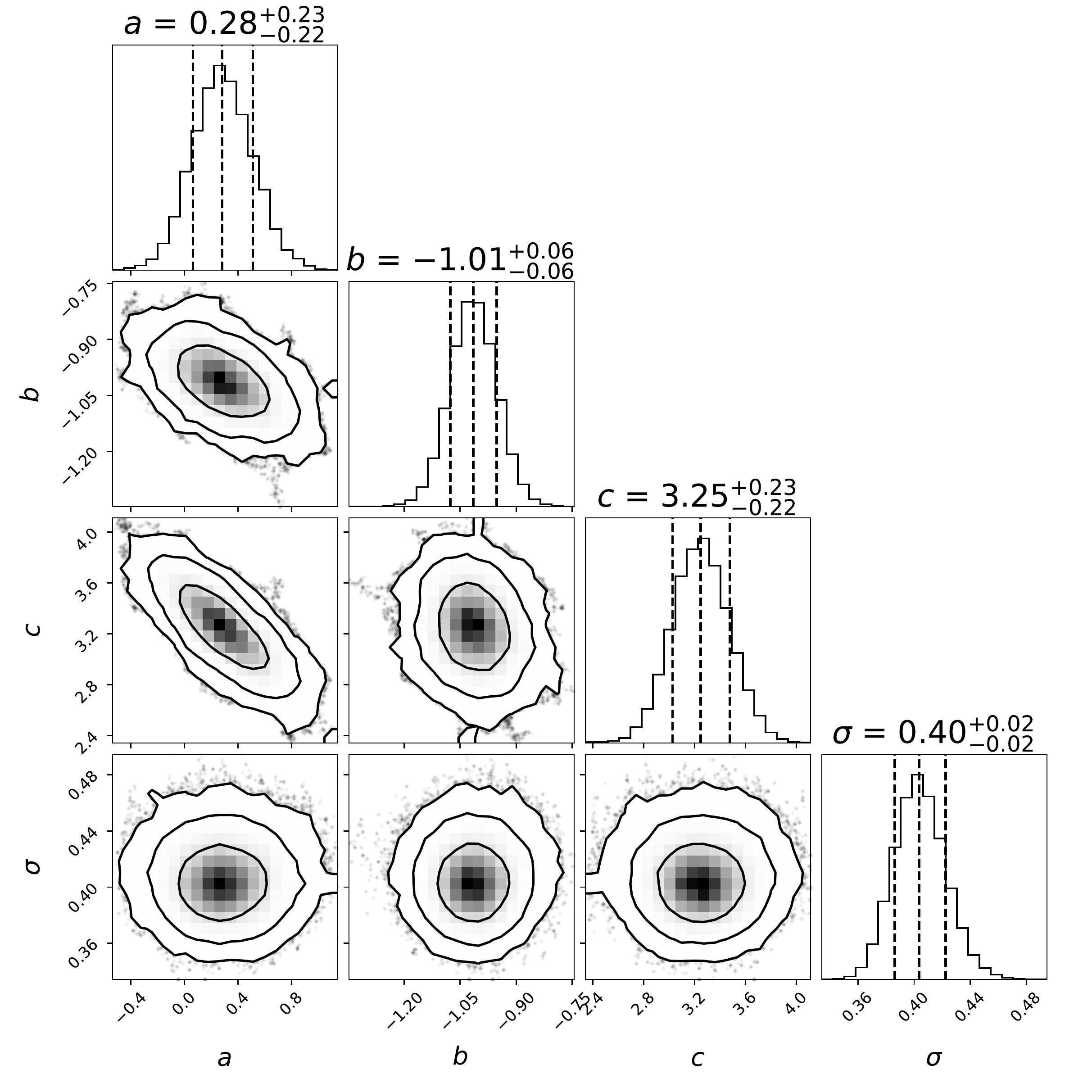}
\caption{Results of the Bayesian analysis applied to the dataset $(log \rho_i, \, \log D_i, \, \alpha_i)$, with $\log\!\rho$ as the dependent variable. The panels show the marginalized PDFs of the model parameters $\boldsymbol{\theta} = (a,b,c;\sigma)$.  Contours correspond to the 68.3, 95.4, and 99.7\% confidence levels, respectively. The median values and their marginalized $1\sigma$ uncertainties are denoted by dashed lines on the histograms showing the posterior probability distributions of the model parameters.}
\label{fig:PDFplane}
\end{figure}

\begin{figure}[!t]
\centering
\vspace{0.75cm}
\includegraphics[width=0.8\textwidth]{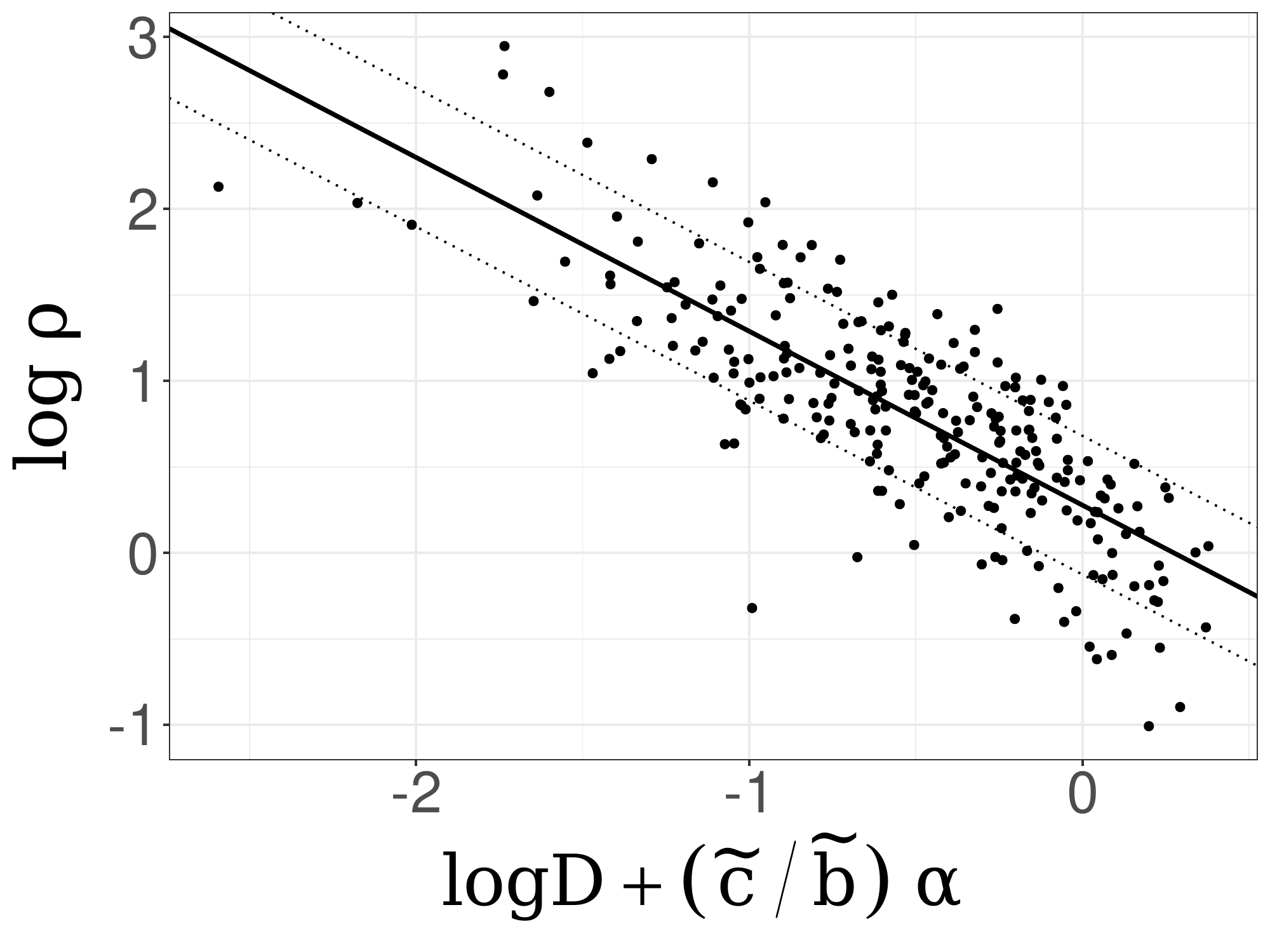}
    \caption{Data points from the set $(log \rho_i, \, \log D_i, \, \alpha_i)$ plotted on the projected plane. The Solid line corresponds to the regression line $\log\,\!\rho=\tilde{a}+\tilde{b}\,\log\!D+\tilde{c}\,\alpha$, while the dashed lines indicate the $\pm \tilde{\sigma}$ deviation of the relation; see Table\,\ref{tab:results} for the median-fit regression parameters' values $(\tilde{a},\tilde{b},\tilde{c};\tilde{\sigma}$).}
\label{fig:plane}
\end{figure}

In this approach, our statistical model $\mathcal{M}$ states that the response variable is sampled from the normal distribution $y \sim \mathcal{N}\!\left(\overline{y};\,\sigma\right)$ with mean
\begin{equation}
\overline{ \log\!\rho} = a + b \, \log\!D + c \, \alpha
\end{equation}
and variance $\sigma$ representing the intrinsic spread related to the unspecified variables. With this approach, we aim to determine, the posterior probability density function (PDF) of the model parameters $\boldsymbol{\theta} = (a,b,c;\sigma)$ via the Bayes' theorem
\begin{equation}
P\!\left(\boldsymbol{\theta}|\mathcal{D}, \mathcal{M}\right) = \frac{P\!\left(\mathcal{D}|\boldsymbol{\theta},\mathcal{M}\right) \, P\!\left(\boldsymbol{\theta}|\mathcal{M}\right)}{P\!\left(\mathcal{D}|\mathcal{M}\right)} \, ,
\end{equation}
where $P\!(\mathcal{D}|\boldsymbol{\theta},\mathcal{M})$ is the likelihood, i.e. the probability of measuring the set of data $\mathcal{D}$ given particular values of the model parameters $\boldsymbol{\theta}$,  $P\!(\boldsymbol{\theta}|\mathcal{M})$ is the prior, i.e. the probability distribution for the parameters in the model, and finally
\begin{equation}
P\!\left(\mathcal{D}|\mathcal{M}\right)=\int P\!\left(\mathcal{D}|\boldsymbol{\theta},\mathcal{M}\right) \, P\!\left(\boldsymbol{\theta}|\mathcal{M}\right) \, d\boldsymbol{\theta}
\end{equation}
is the model evidence.

We perform the analysis following closely \citet{Ostorero17}, in particular using the code APEMoST\footnote{Automated Parameter Estimation and Model Selection Toolkit; \url{ http://apemost.sourceforge.net/}, 2011 February.}  \citep{Gruberbauer09} with $2\times 10^6$ MCMC iterations and twenty chains to ensure a sufficiently complete sampling of the parameter space. We adopt independent, non-informative uniform priors on $(a,b,c)$, with the parameter space boundaries set to  $[-10,10]$. For the intrinsic spread $\sigma$, on the other hand, which by definition is always a positive number, we assume the distribution function that describes a variate with mean $r/\mu$, and variance $r/\mu^2$, namely
\begin{equation}
P\!\left(\sigma|\mathcal{M}\right)=\frac{\mu^r}{\Gamma(r)} \, x^{r-1} \, \exp(-\mu x) \, ,
\end{equation}
where $x= 1/\sigma$, and $\Gamma(r)$ is the Euler gamma function; in our calculations, we set $r=\mu=10^{-5}$ and the variability interval boundaries $[0.01,1000]$. The random number generator was set with bash command \texttt{GSL\_RNG\_TYPE="taus"} and the initial seed of the random number generator was set with \texttt{GSL\_RNG\_SEED=\$RANDOM}.

\begin{figure}[!t]
\centering
\includegraphics[width=0.8\textwidth]{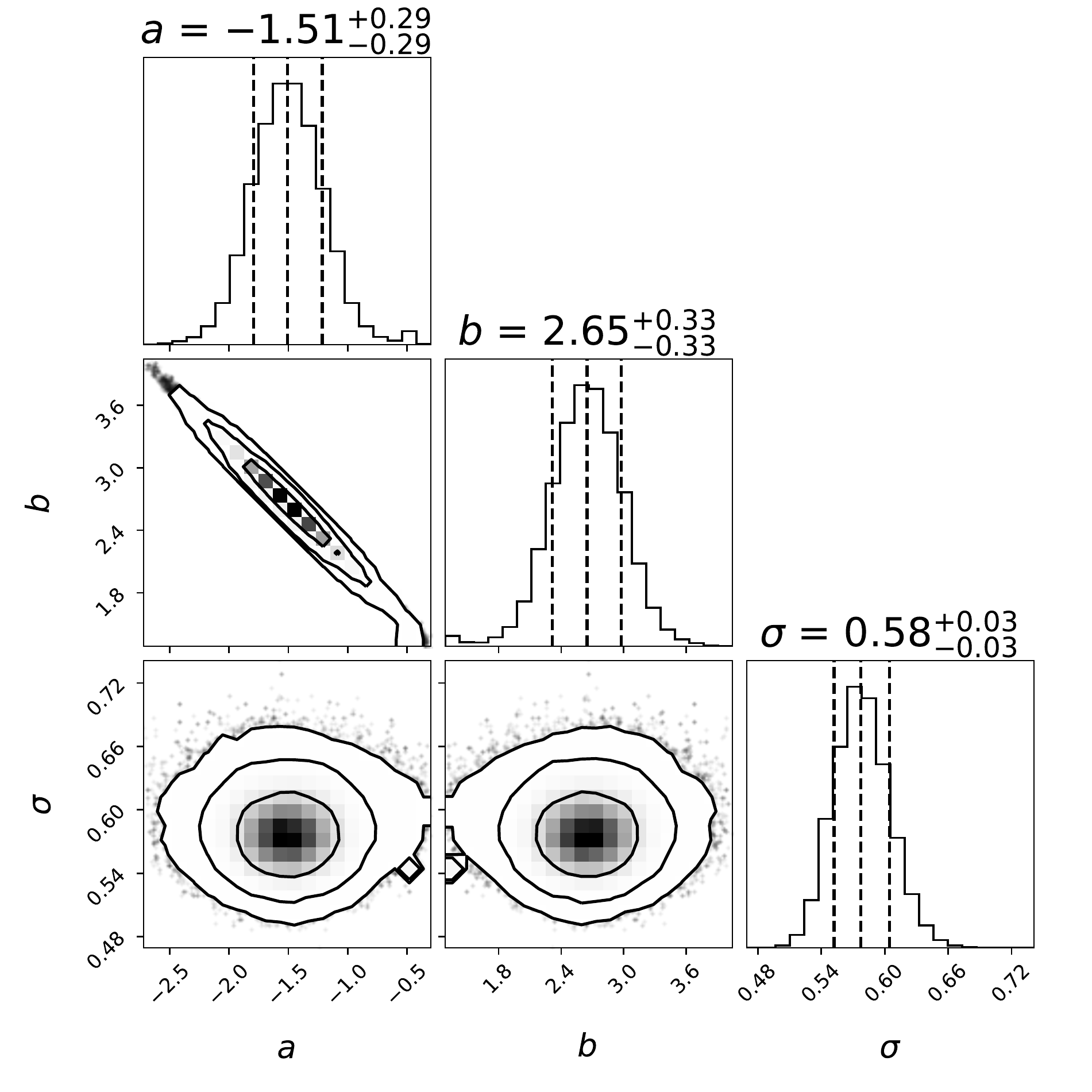}
  \caption{Results of the Bayesian analysis applied to the dataset $(log \rho_i, \, \alpha_i)$, with $\log\!\rho$ as the dependent variable. The panels show the marginalized PDFs of the model parameters $\boldsymbol{\theta} = (a,b;\sigma)$.  Contours correspond to the 68.3, 95.4, and 99.7\% confidence levels, respectively. The median values and their marginalized $1\sigma$ uncertainties are denoted by dashed lines on the histograms showing the posterior probability distributions of the model parameters.}
  \label{fig:2D-PDF}
\end{figure}

The results of the Bayesian regression analysis as described above, are summarized in Table\,\ref{tab:results}, and presented in Figure\,\ref{fig:PDFplane}. In order to better visualize the significance of the correlation, as well as spread in the data around the correlation plane, in Figure\,\ref{fig:plane} we plot the data points from the set $\mathcal{D}$ along the projected plane with the median-fit correlation parameters' values $(\tilde{a},\tilde{b},\tilde{c})$ and $\pm \tilde{\sigma}$ deviations (see Table\,\ref{tab:results} for the parameter values). We see that for any pair of values of $\log\!D$ and $\alpha$, the density of the ambient medium $\log\!\rho$ falls within a factor $\simeq 0.80$ from the mean relation at $68\%$ confidence level.

\begin{figure}[!t]
\centering
\vspace{0.75cm}
    \includegraphics[width=0.8\textwidth]{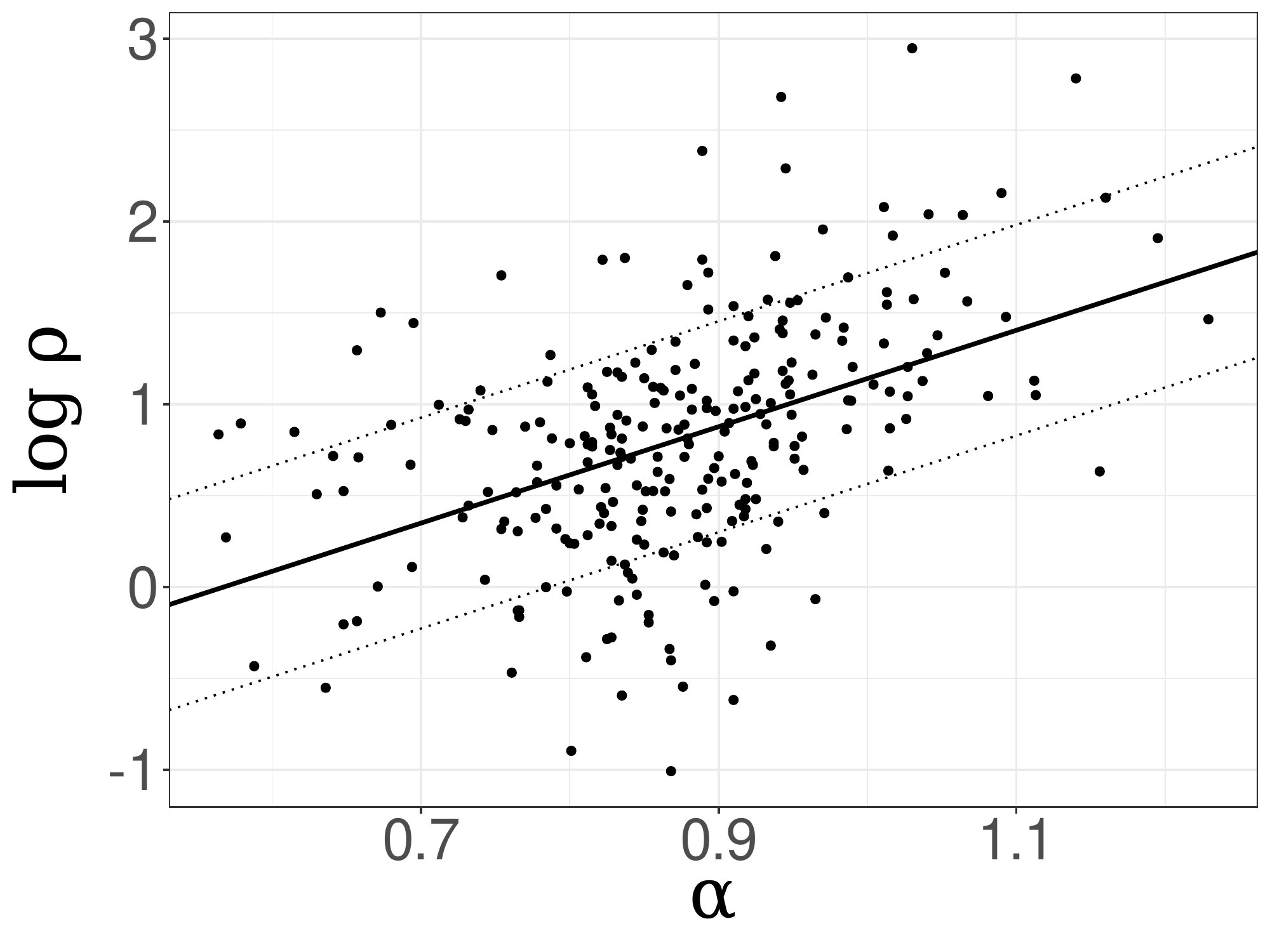}
    \caption{Data points from the set $(log \rho_i, \, \alpha_i)$ plotted with the linear regression line $\log\,\!\rho=\tilde{a}+\tilde{b}\,\alpha$ (solid line), with $\pm \tilde{\sigma}$ deviations (dashed lines); see Table\,\ref{tab:results} for the median-fit regression parameters' values $(\tilde{a},\tilde{b};\tilde{\sigma}$).}
    \label{fig:2D}
    \end{figure}    

\subsection{Simplified Linear Relation}
\label{sec:simplified}

Since it is not always possible to measure the projected linear sizes of distant radio galaxies, we also investigate whether spectral information alone, i.e., the slope of the observed radio emission continuum between two selected frequencies, is sufficient to estimate the density of the ambient medium on intergalactic scales. In our "reduced'' training dataset $(\log \rho_i, \, \alpha_i)$, we still see a statistically significant correlation between $\log\!\rho$ and $\alpha$, for which Pearson's product-moment correlation test returns $\rho=0.459$, and Kendall's rank correlation gives $\tau=0.328$; the $p$-values are $< 10^{-14}$ in both tests.

We analyze this correlation further in a Bayesian approach, in the same manner as outlined in Sections\,\ref{sec:plane} assuming in particular that our response variable $\log\!\rho$ is sampled from the normal distribution $\log\!\rho \sim \mathcal{N}\!\left(a + b \, \alpha;\,\sigma\right)$, and seeking the posterior PDF of the model parameters $\boldsymbol{\theta} = (a,b;\sigma)$. The results of the analysis are summarized in Table\,\ref{tab:results}, and visualized in Figures\,\ref{fig:2D-PDF} and \ref{fig:2D}. For any given value of the spectral index $\alpha$, the density of the ambient medium, $\log\!\rho$, falls within a factor  $\simeq 1.16$ from the mean relation at the 68\% confidence level.

\begin{figure}[!t]
\centering
\includegraphics[width=0.8\textwidth]{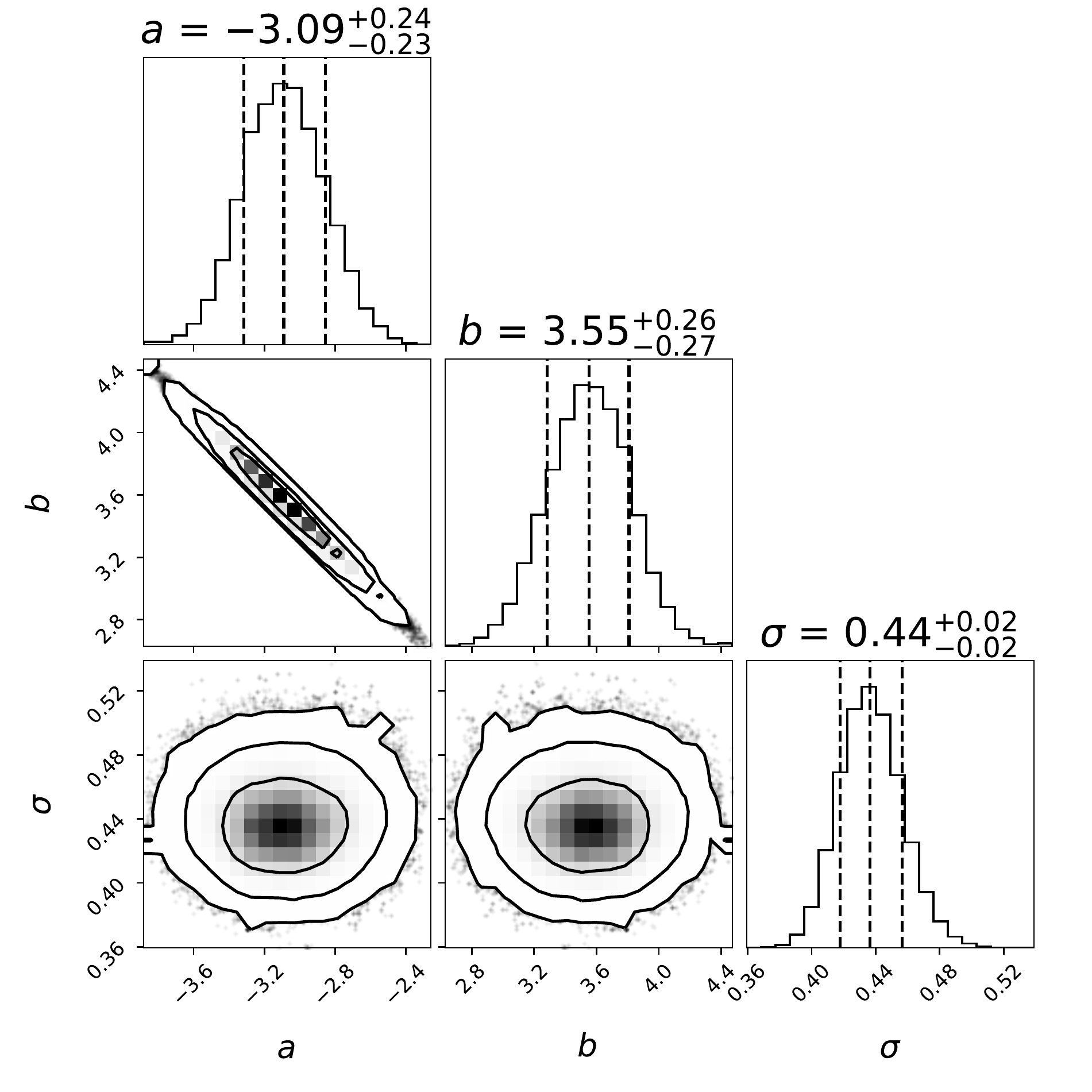}
  \caption{Results of the Bayesian analysis applied to the dataset $(\log\![\rho_0 a_0^{3/2}]_i, \, \alpha_i)$, with $\log\!\rho_0$ as the dependent variable. The panels show the marginalized PDFs of the model parameters $\boldsymbol{\theta} = (a,b;\sigma)$. Contours correspond to the 68.3, 95.4, and 99.7\% confidence levels, respectively. The median values and their marginalized $1\sigma$ uncertainties are denoted by dashed lines on the histograms showing the probability distributions of the model parameters.}
  \label{fig:2Drho0-PDF}
\end{figure}

\begin{figure}[!t]
\centering
\vspace{0.75cm}
    \includegraphics[width=0.8\textwidth]{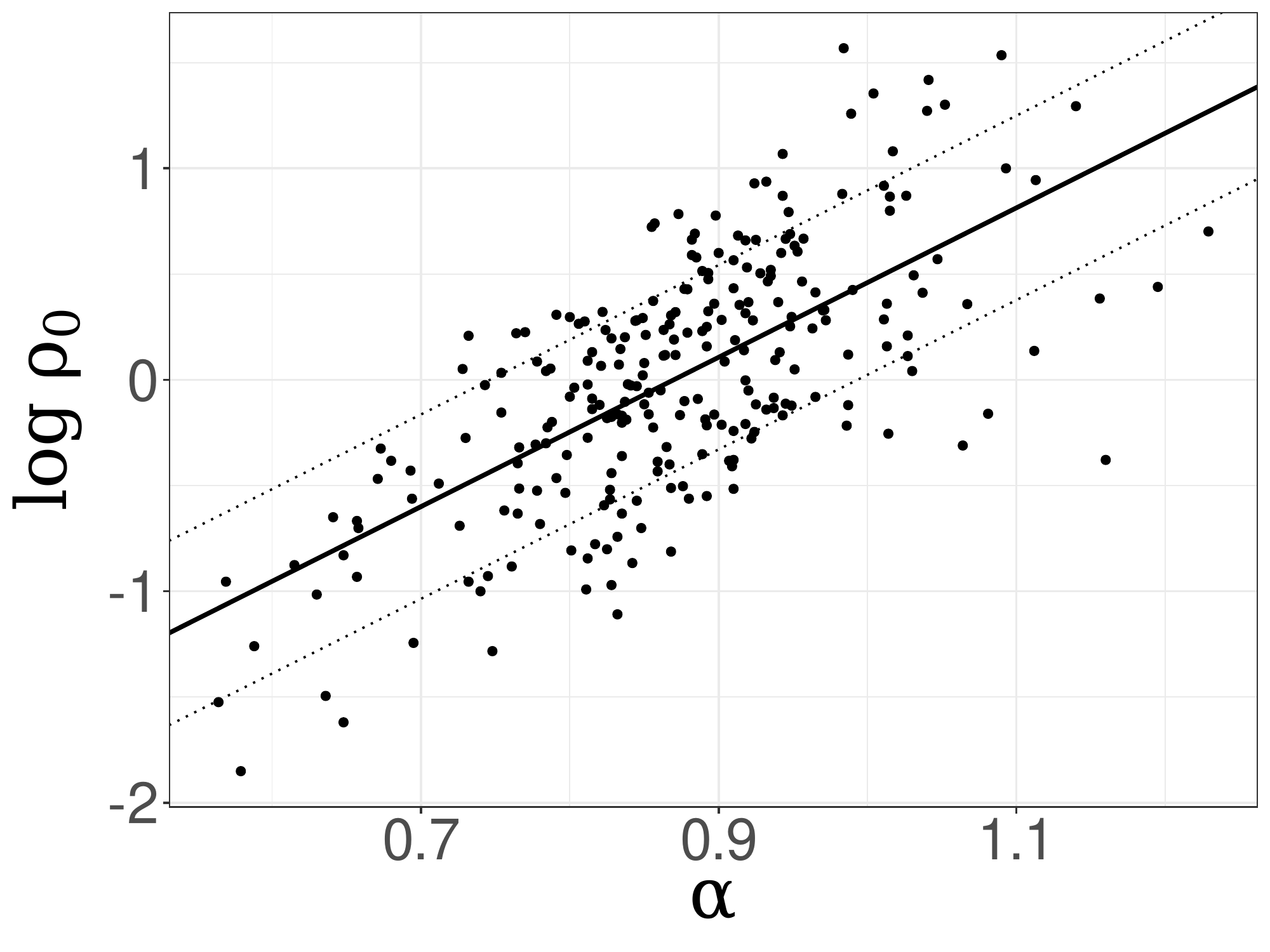}
    \caption{Data points from the set $(\log\![\rho_0 a_0^{3/2}]_i, \, \alpha_i)$ plotted with the linear regression line $\log\![\rho_0 a_0^{3/2}]=\tilde{a}+\tilde{b}\,\alpha$ (solid line), with $\pm \tilde{\sigma}$ deviations (dashed lines); see Table\,\ref{tab:results} for the median-fit regression parameters' values $(\tilde{a},\tilde{b};\tilde{\sigma}$).}
    \label{fig:2Drho0}
    \end{figure}

\subsection{Alternative Relations}  
\label{sec:alternative}

The relatively large spread in the simplified $\log\!\rho \propto \alpha$ correlation, is predominantly due to the fact that the variable $\log\!\rho$ describes the density of the gas just ahead of the jets' termination shocks, i.e. at the distances $r \sim D/2$ from the centers of the systems; with the assumed density profile $\rho(r) \propto r^{-3/2}$ (Equation\,\ref{eq:rho}), this variable has to be therefore a rather strong function of the source linear size $D$, which itself varies rather substantially between the objects included in our sample (in particular, from several kpc up to 1\,Mpc). Any correlation not explicitly involving the parameter $\log\!D$, must therefore be characterized by a large scatter. 

In order to minimize this dependence, and in this way, to narrow the spread in the simplified linear relation involving solely the radio spectral information, we replace the response variable $\log\!\rho$ with $\log\!\rho_0 a_0^{3/2}$. The units for the latter are $10^8$\,cgs, so that for example with $a_0=10$\,kpc, the $\log\!\rho_0 a_0^{3/2} = 0$ value corresponds to the gas number density $n_0\simeq \rho_0/m_p \simeq 10^{-2}$\,cm$^{-3}$.

First, however, we analyze the modified correlation plane, by forming an ``alternative'' training dataset $(\log\![\rho_0 a_0^{3/2}]_i, \, \log D_i, \, \alpha_i)$. We repeat the correlation analysis and the Bayesian regression analysis for this dataset, results of which are summarized in Table\,\ref{tab:results}. As given, the change in the response variable does not dramatically alter the correlation significance, or the spread inferred from the Bayesian analysis, as long as both prediction variables $\log\!D$ and $\alpha$ are included: the overall correlation is similarly very significant ($p$-values $< 2.2 \times 10^{-16}$ for both tests, with the Pearson's test yielding $\rho=0.757$, and Kendall's rank  correlation yielding $\tau=0.520$), while the response variable $\log\!\rho_0 a_0^{3/2}$ falls within a factor $\simeq 0.78$ from the mean relation at the $68\%$ confidence level for any given pair $(\log\!D, \alpha)$.

Finally, we analyze the ``alternative reduced'' training dataset $(\log\![\rho_0 a_0^{3/2}]_i, \, \alpha_i)$, in which we do see an improved correlation between $\log \rho_0 a_0^{3/2}$ and $\alpha$ (compared to that seen between $\log\!\rho$ and $\alpha$),  with the Pearson's product-moment correlation test yielding $\rho=0.675$ and the Kendall's rank correlation yielding $\tau=0.452$ with $p$-values $< 2.2 \times 10^{-16}$ in both tests. We repeat the Bayesian regression analysis also for this dataset; the results are summarized in Table\,\ref{tab:results}, and visualized in Figures\,\ref{fig:2Drho0-PDF} and \ref{fig:2Drho0}. This time, for every value of the spectral index $\alpha$, the \emph{central} density of the ambient medium $\log \rho_0 a_0^{3/2}$ falls within a factor $\simeq 0.88$ from the mean relation at the 68\% confidence level.

\section{Discussion}
\label{sec:discussion}

As mentioned in the Introduction, the direct method of estimating the density of the hot gaseous fraction of the IGM, relies on detailed modeling of good-quality X-ray measurements of the extended, thermal emission component surrounding radio structures. Such data are currently available only for a small fraction of the nearby and/or brightest sources subjected to deep exposures with either the Chandra or XMM-Newton X-ray telescopes. 

In this context, first we refer to the results presented in \citet{Belsole07}, who analyzed the available Chandra and XMM-Newton data for powerful FR\,IIs selected from the Third Cambridge Catalogue of Radio Sources, Revised Edition \citep[3CRR;][]{Laing83}, within the redshift range $0.45 < z < 1.0$, and with the 178\,MHz power spectral density between $\sim 7 \times 10^{26}$\,W\,Hz$^{-1}$\,sr$^{-1}$ and $\sim 10^{28}$\,W\,Hz$^{-1}$\,sr$^{-1}$ (cf. Figure\,\ref{fig:P-z} in Appendix\,\ref{ab} for the distribution of the 1.4\,GHz power with redshift for the sources included in our sample). 

\begin{figure}[!t]
\centering
    \includegraphics[width=0.48\textwidth]{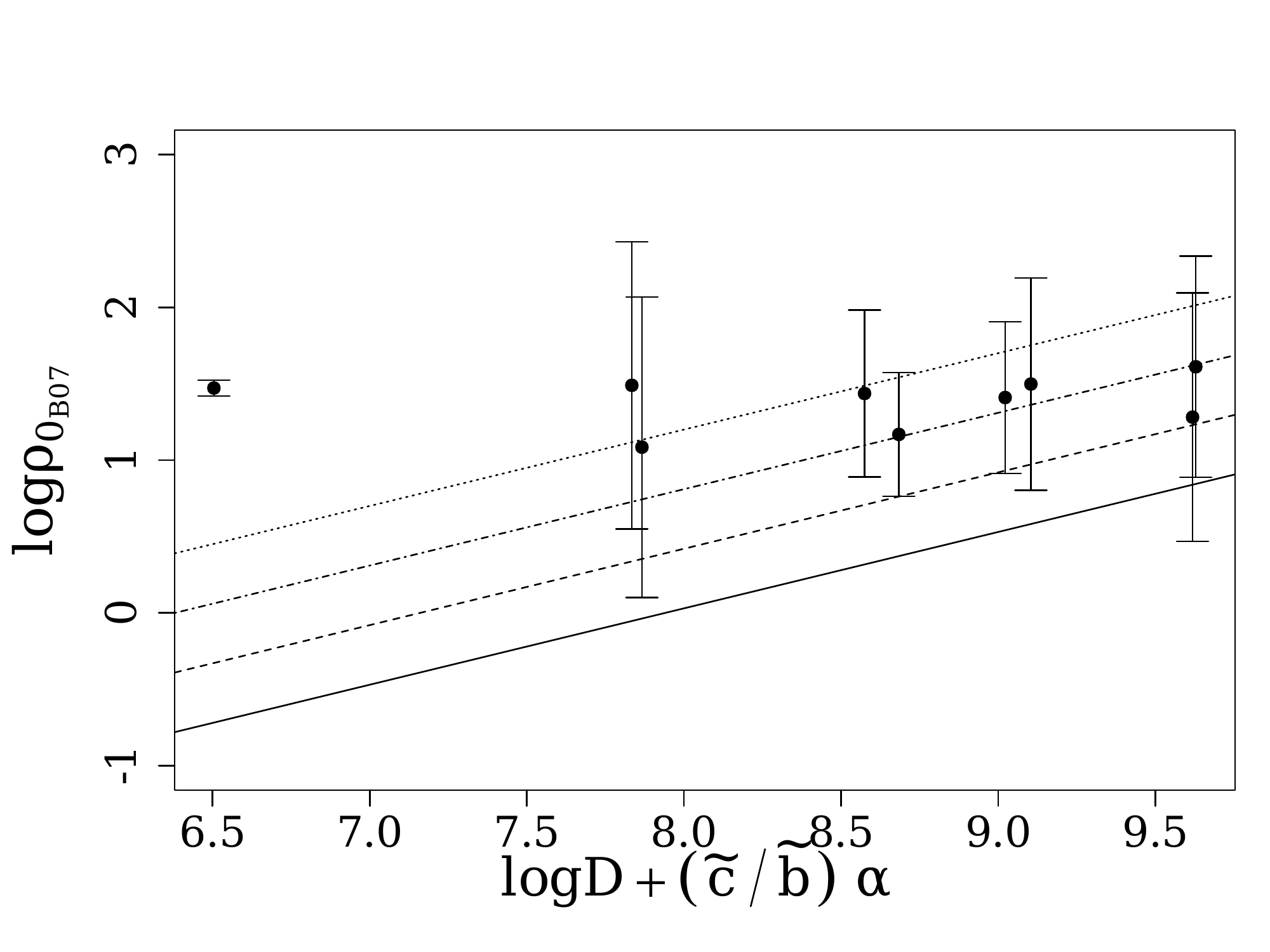}
    \includegraphics[width=0.48\textwidth]{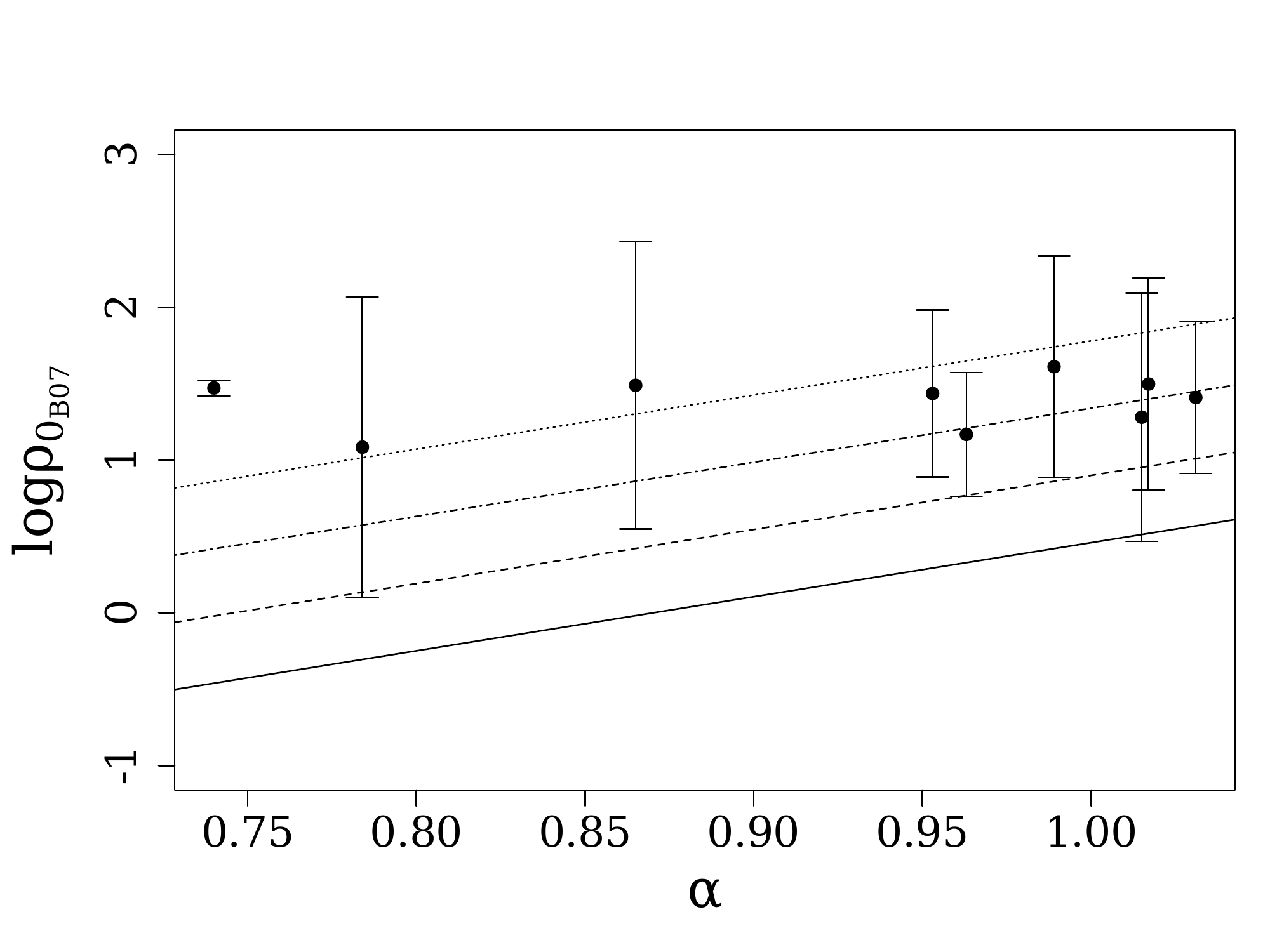}
    \caption{Comparison between the parameter $\log \rho_0 r_c^{3/2}$ [ $10^8$\,g\,cm$^{-3/2}$ units], calculated based on the best-fit values for $n_0$ and $r_c$ obtained by \citet{Belsole07} from the X-ray data analysis, 
    and the regression (solid) lines following from our Bayesian analysis assuming the model $\log \rho_0 a_0^{3/2} \sim \mathcal{N}\!\left(a + b \, \log\!D + c \, \alpha,\sigma\right)$, or the $\log \rho_0 a_0^{3/2} \sim \mathcal{N}\!\left(a + b \, \alpha,\sigma\right)$ (left and right panels, respectively; see Section\,\ref{sec:alternative} and Table\,\ref{tab:results}). Dashed, dot-dashed, and dotted lines correspond to $+1\sigma$, $+2\sigma$, and $+3\sigma$ intervals from the best-fit relations, respectively.}
    \label{fig:B07}
\end{figure}

Assuming the X-ray surface brightness profiles, corresponding to the IGM emission component, in a form $\Sigma_{\rm X} \propto [1+(r/r_c)^2]^{-3\tilde{\beta}+0.5}$, \citet{Belsole07} were able to constrain the parameters $\tilde{\beta}$ and $r_c$ for 10 out of the selected 20 targets. We note that with such a parameterization of the surface brightness, the corresponding gas density profile reduces to the one given in our equation\,\ref{eq:rho} for $\beta = 3\tilde{\beta}$ and $r_c = a_0 \ll r$. Nine of those 10 sources overlap with our sample, namely 3C\,200, 3C\,207, 3C\,220.1, 3C\,254, 3C\,265, 3C\,292, 3C\,295, 3C\,330, and 3C\,427.1. For these, based on the best-fit values provided in \citeauthor{Belsole07} for the central gas number density $n_0$ and the core radius $r_c$, we therefore calculate the parameters $\log \rho_0 r_c^{3/2}$ with $\rho_0 = m_p n_0$, which can be confronted directly with the regression lines following from our Bayesian regression analysis, keeping in mind that in our modelling $a_0$ and $\rho_0$ are not independent parameters, as discussed in Section\,\ref{sec:alternative}. The results of this comparison are presented in Figure\,\ref{fig:B07}. 

As shown, the central gas densities emerging from the observational constraints by \citet{Belsole07}, are within $3\sigma$ from our best-fit relations, for all the sources but 3C\,295 (given the errors). These ``X-ray'' values are, nonetheless, in all the cases systematically larger than the ones implied by our modeling and regression analysis. The reasons for such a discrepancy can be, at least partially, the difference between the universal $\beta$ parameter assumed in the DYNAGE modeling, and the $\tilde{\beta}$ values following from the X-ray data fitting: indeed, even though the best-fit values of $\tilde{\beta}$ provided in \citet{Belsole07} are in many cases consistent within the errors with the ``universal'' $\tilde{\beta}=0.5$, they are nonetheless in the range $0.5 \lesssim  \tilde{\beta} \lesssim 1.0$, implying steeper slopes of the ambient density profiles than the one assumed in the DYNAGE modeling. 

We also observed that, for all the overlapping sources except 3C\,295, the observational constraints on $r_c$ are particularly uncertain, with the relative errors exceeding $50\%$. This is due to the extended wings of the point spread functions of the bright unresolved X-ray cores of the targets which, even with the superb ($\sim$\,arcsec) angular resolution of the Chandra ACIS imaging instrument, precludes any precise determination of the diameter of a central plateau in their hot gaseous atmospheres (which is typically of the order of a few/several arcsec, or less). Importantly, the source 3C\,295, for which the observational constraints on $r_c$ are relatively tight, i.e. $r_c \simeq 3.4\pm 0.25$\,arcsec, is also the one characterized by the largest discrepancy between the two values of the central gas density. This particular object is, on the other hand, the smallest radio galaxy in this overlapping sub-sample, with the linear size of its radio structure $D \simeq 34$\,kpc; and since $D/2 \lesssim r_c \simeq 18.8 \pm 1.4$\,kpc, we conclude that the DYNAGE model assumption, regarding a self-similar evolution of the radio structure in a power-law ambient density profile, is hardly fulfilled for the target. This signals that the method proposed here, may not apply to compact radio galaxies.

Meanwhile, for the majority of sources for which the X-ray surface brightness profiles cannot be constrained at all (or cannot be constrained precisely) from the X-ray imaging data, but the X-ray emission of the IGM can still be seen in the total radiative outputs of the systems, one can consider the total IGM-related X-ray luminosity as a proxy for the central gas density. Indeed, considering solely the free-free continuum component of the thermal IGM emission, $\beta=3/2$ gas density profile, and uniform gas temperature within the entire emission volume, this luminosity reads as
\begin{eqnarray}
L_{\rm X} &  \propto & \int \rho^2 dV \propto  \int_0^{a_0} \rho_0^2  \, r^2 dr + \int_{a_0}^{r_t}  \left[\rho_0 a_0^{3/2}\right]^2 r^{-1} dr  \nonumber \\
&&  \sim  \left[\rho_0 a_0^{3/2}\right]^2 \times \left(\frac{1}{3}+\ln \frac{r_t}{a_0}\right) \, , 
\label{eq:Xrho}
\end{eqnarray}
meaning an approximate square dependence of $L_{\rm X}$ on $\rho_0 a_0^{3/2}$, modulo the slowly varying factor $(\frac{1}{3}+\ln r_t/a_0)$ for the termination radius $r_t \gg a_0$.

\begin{figure}[!t]
\centering
    \includegraphics[width=0.48\textwidth]{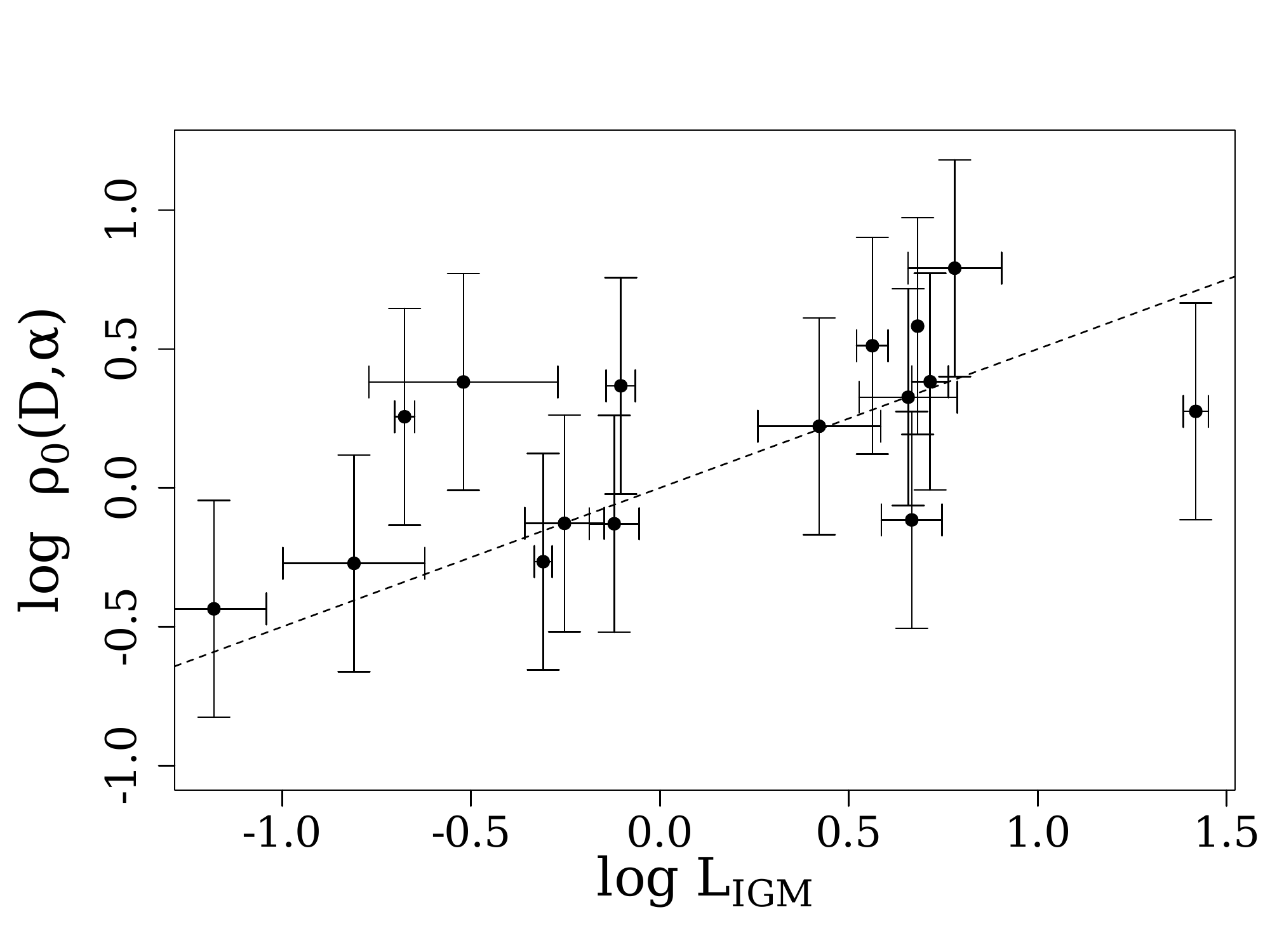}
    \includegraphics[width=0.48\textwidth]{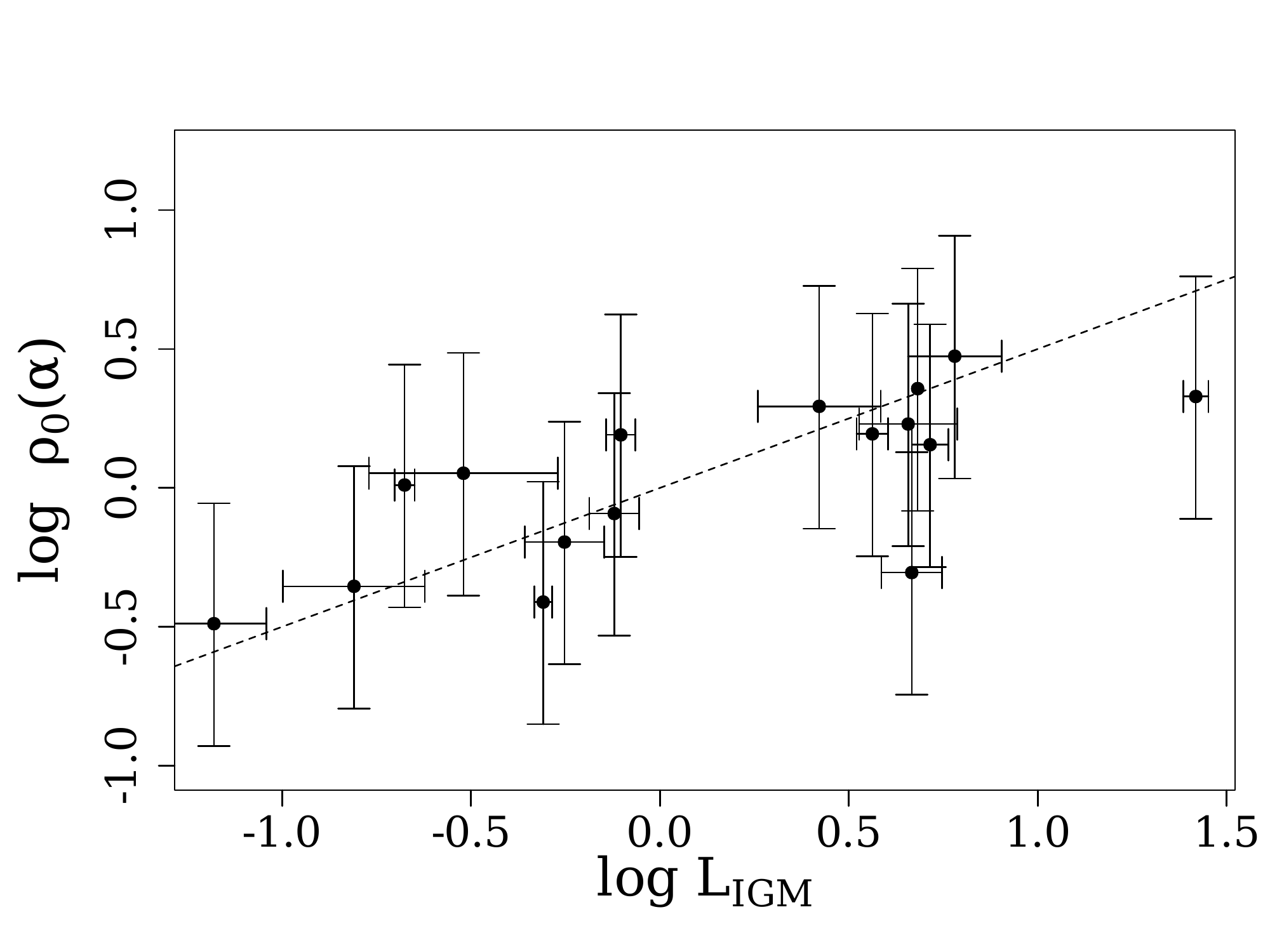}
    \caption{Comparison between the IGM-related total X-ray luminosities (within the $r_{500}$ radii), in the units of $10^{43}$\,erg\,s$^{-1}$, as estimated by \citet{Ineson17}, and the $\log \rho_0 a_0^{3/2} \sim \mathcal{N}\!\left(a + b \, \log\!D + c \, \alpha,\sigma\right)$, or the $\log \rho_0 a_0^{3/2} \sim \mathcal{N}\!\left(a + b \, \alpha,\sigma\right)$ values (left and right panel, respectively), following from our Bayesian regression analysis, as discussed in Section\,\ref{sec:alternative} (see also Table\,\ref{tab:results}). The dashed lines represent the $y=\frac{1}{2}x$ scaling, that can be expected based on equation\,\ref{eq:Xrho}, included to guide the eye.}
    \label{fig:I17}
\end{figure}

Keeping the above relation in mind thereafter we refer to the results presented in \citet{Ineson17}, who studied a sample of FR\,II radio galaxies with available good X-ray coverage, selected from various flux-limited, spectroscopically complete, low-frequency radio surveys. The sample is representative for radio-loud active galaxies at redshifts $z\sim 0.1$ and $z\sim 0.5$, and contains the objects with the 151\,MHz power spectral density between $\sim 0.2 \times 10^{26}$\,W\,Hz$^{-1}$\,sr$^{-1}$ and $0.4 \times 10^{28}$\,W\,Hz$^{-1}$\,sr$^{-1}$. For these objects, \citet{Ineson17} estimated the IGM-related total X-ray luminosities (within the $r_{500}$ radii), $L_{\rm IGM}$, which are plotted in Figure\,\ref{fig:I17} for the objects overlapping with our list against the $\log \rho_0 a_0^{3/2} \sim \mathcal{N}\!\left(a + b \, \log\!D + c \, \alpha,\sigma\right)$, or the $\log \rho_0 a_0^{3/2} \sim \mathcal{N}\!\left(a + b \, \alpha,\sigma\right)$ values, following from our Bayesian regression analysis. 

There is a positive, statistically significant correlation between $L_{\rm IGM}$ and $\log \rho_0 a_0^{3/2}$. In particular, for the $\log \rho_0 a_0^{3/2} \sim \mathcal{N}\!\left(a + b \, \log\!D + c \, \alpha,\sigma\right)$ model the Pearson's product-moment correlation test yields $\rho=0.613$ and $p$-value $\simeq 0.012$, and the Kendall's rank correlation yields $\tau=0.517$ and $p$-value $\simeq 0.005$. Meanwhile, the $\log \rho_0 a_0^{3/2} \sim \mathcal{N}\!\left(a + b \, \alpha,\sigma\right)$ model seems to work equally well in this respect, with the corresponding parameters $\rho=0.716$ and $p$-value $\simeq 0.002$ for  the Pearson's test, and $\tau=0.60$ and $p$-value $\simeq 0.001$ for the Kendall's rank correlation.

\section{Conclusions}
\label{sec:conclusions}

Our analysis of the data emerging from the extensive DYNAGE modeling of a large sample of FR\,II radio sources, reveals a positive and statistically significant correlation between the density of the ambient medium at scales comparable to the linear sizes of the sources' radio structures, $\log\!\rho$, and the spectral index between the emitted frequencies 0.4\,GHz and 5\,GHz, $\alpha$. Moreover, the intrinsic scatter in the correlation --- which may be interpreted as a combination of the uncertainty resulting from the DYNAGE model assumptions, and the dependence on additional hidden parameters --- can be further minimized by either including the additional variable, in particular the observed projected size of the radio structure $\log\!D$, and/or using the central gas density $\log\!\rho_0$ as a dependent variable instead of $\log\!\rho a_0^{3/2}$ (for the assumed $\beta=3/2$ density profile); in either of these cases, the statistical model allows us to predict the density of the ambient medium with an accuracy of about $0.3$\,dex.

The fact that there is \emph{some positive} correlation between the environment density of FR\,IIs and the slope of the integrated radio emission of their extended lobes, should not, in fact, be that surprising, keeping in mind the work by \citet{OSullivan15}, who reported on the absence of radio galaxies with a large degree of linear polarization (at 1.4\,GHz) and steep radio spectra. When combined with the analysis of the neighbouring galaxy counts, this observational finding indicates that the decreased polarization degree correlates with the richness of the local environment (as expected if the depolarization is predominantly due to the Faraday effect related to the surrounding magnetoionic medium), and hence that the spectral index is indeed positively correlated with the environment density (see the discussion in Section 3.4.4. therein).

There are several important implications of the emerging dependence, each deserving an in-depth discussion. One is related to the physics of hotspots in FR\,II radio galaxies, which constitute spectacular manifestations of mildly-relativistic shocks formed when extremely low-density but relativistic and highly magnetized outflows (jets), collide with high-density but non-relativistic gas of the IGM \citep[e.g.,][]{Meisenheimer89}, and in particular to the issue of particle acceleration taking place at and around the fronts of such shocks. Namely, keeping in mind the $\log\!\rho - \alpha$ positive relation on one hand, and the $\alpha - \alpha_{\rm inj}$ tight and also positive correlation on the other hand, one may ask how exactly could the increasing density of the ambient medium decrease the efficiency of accelerating jet electrons at the termination shocks to ultra-relativistic energies.

The other emerging ramification of this study is the general problem of high-redshift radio galaxies \citep[for a review see][]{McCarthy93,Miley08}, which are known to display steeper radio emission continua than their low-redshift counterparts \citep[e.g.,][]{Klamer06}. Those topics will be addressed in following publications. Here, instead, we only report on the newly recognized correlations, and on how our findings seem to be supported by the analysis of good-quality and high-angular resolution X-ray data available for some of the targets. In particular, this comparison signals that the main limitation of our model is related to the assumption of a self-similar evolution of a radio structure in a power-law ambient medium density $\rho \propto r^{-3/2}$. As  a result, the method proposed here should be taken with caution when dealing with the most compact targets, i.e. those with $D \sim \mathcal{O}(10\,{\rm kpc})$ and smaller, and also with giants, for which $D \sim \mathcal{O}(1\,{\rm Mpc})$ .

With proper multi-wavelength support, however, one could possibly improve the emerging statistical correlations, for the purpose of using them as well-established cosmological tools when dealing with large samples of objects emerging from wide-area and high-sensitivity radio surveys \citep[such as the one carried out by LOFAR, or the Square Kilometre Array, or the Very Large Array Sky Survey; e.g.,][respectively]{Hardcastle19,Agudo15,Lacy20}, combined with the modern and next-generation optical photometric and spectroscopic surveys \citep[like the Sloan Digital Sky Survey, or the forthcoming Large Synoptic Survey Telescope;][respectively]{SDSS,Ivezic19}, and the new-generation X-ray surveys \citep[in particular, eROSITA onboard the Spectrum-Roentgen-Gamma mission; e.g.,][]{Predehl21}.

\begin{acknowledgments}

 We acknowledge stimulating discussions on Bayesian analysis with A. Diaferio. The authors acknowledge also C.~C.~Cheung and S.~P.~O'Sullivan for the discussions and helpful comments on the manuscript.\\
 AW and {\L}S were supported by the Polish National Science Centre grant 2016/22/E/ST9/00061. JM was partly supported by the National Science Centre, Poland grant UMO 2018/29/B/ST9/01793. 
LO acknowledges partial support from the Italian Ministry of Education, University and Research (MIUR) under the Departments of Excellence grant L.232/2016, and from the INFN grant InDark.

\end{acknowledgments}

\appendix

\section{The Spectral Index Correlation} 
\label{aa}

\begin{figure}[H]
\centering
\includegraphics[width=0.8\textwidth]{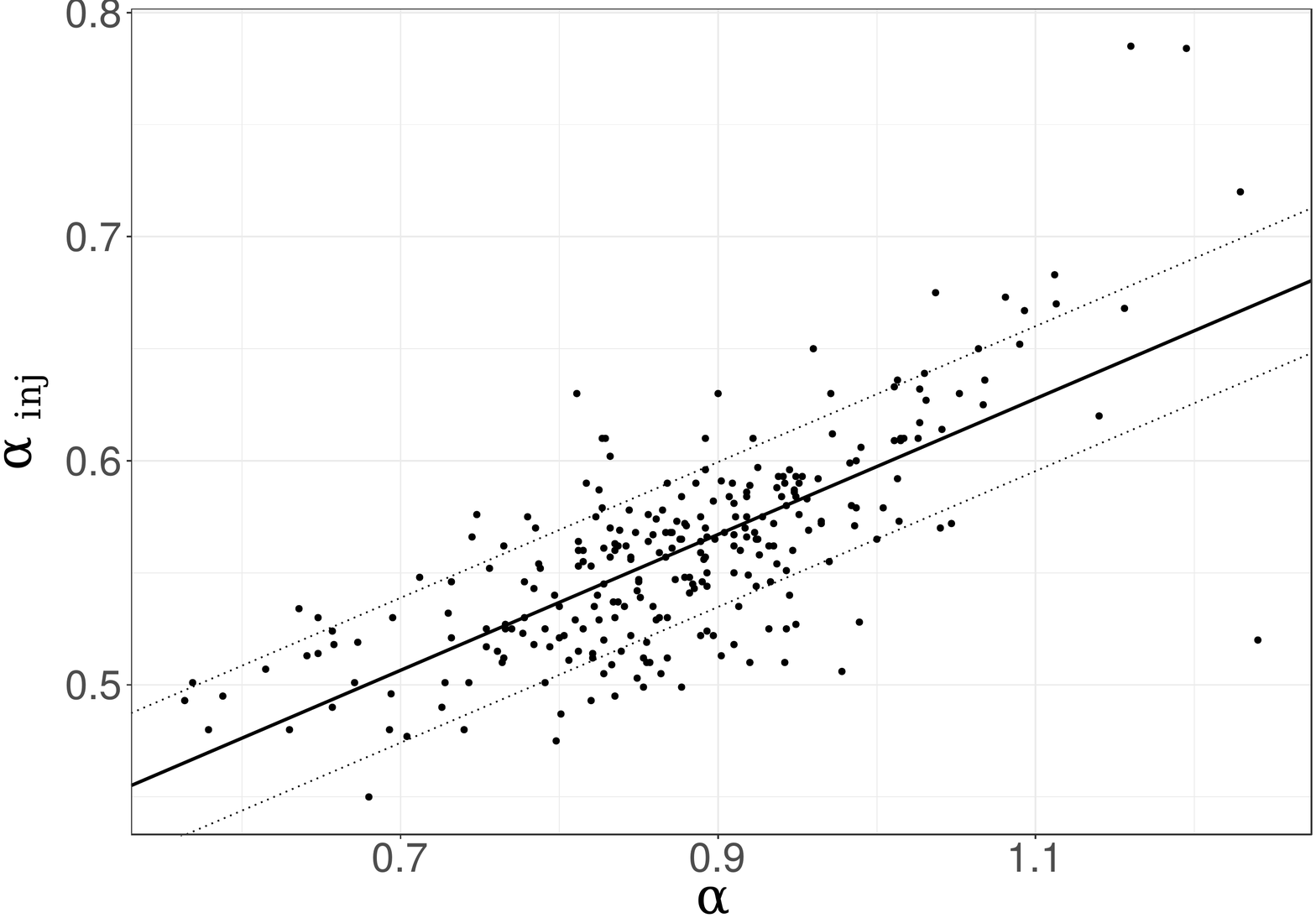}
    \caption{Data points from our DYNAGE training dataset plotted along the correlation $\alpha_{\rm inj} = \tilde{a}+\tilde{b}\,\alpha \pm \sigma$, for the median-fit correlation parameters' values $\tilde{a}=0.294$, $\tilde{b}=0.303$, and $\tilde{\sigma} =0.0322$.}
    \label{fig:alfy}
\end{figure}

As discussed in Section\,\ref{sec:method}, based on the $\rho =f\!(D, \alpha, \alpha_{\rm inj})$ relation expected in the framework of the DYNAGE modeling, we anticipate a simpler scaling $\rho =f\!(D, \alpha)$ involving only these independent variables $D$ and $\alpha$, which are accessible for direct observations. This anticipation is justified, since in our DYNAGE training dataset we see a strong, positive correlation ($p$-value $<2\times 10^{-16}$) between the spectral index $\alpha$ for the two given emitted frequencies 0.4\,GHz and 5\,GHz, and the injection spectral index $\alpha_{\rm inj}$, with $\rho=0.73$ for the Pearson's product-moment correlation test, and $\tau=0.51$ for the Kendall's rank correlation. The Bayesian analysis performed in the same manner as outlined in Section\,\ref{sec:stat}, assuming the model $\alpha_{\rm inj} \sim \mathcal{N}\!\left(a + b \, \alpha;\,\sigma\right)$ for the dependent variable $\alpha_{\rm inj}$, yields the correlation parameter median-fit values $a=0.294 \pm 0.015$, $b=0.303 \pm 0.017$, and the intrinsic scatter $\sigma = 0.0322^{+0.0015}_{-0.0014}$; this correlation, overlaid with the data points, is displayed in Figure\,\ref{fig:alfy}.

\section{The Sample and the Model Data} 
\label{ab}

\begin{figure}[t!]
\centering
\includegraphics[width=0.8\textwidth]{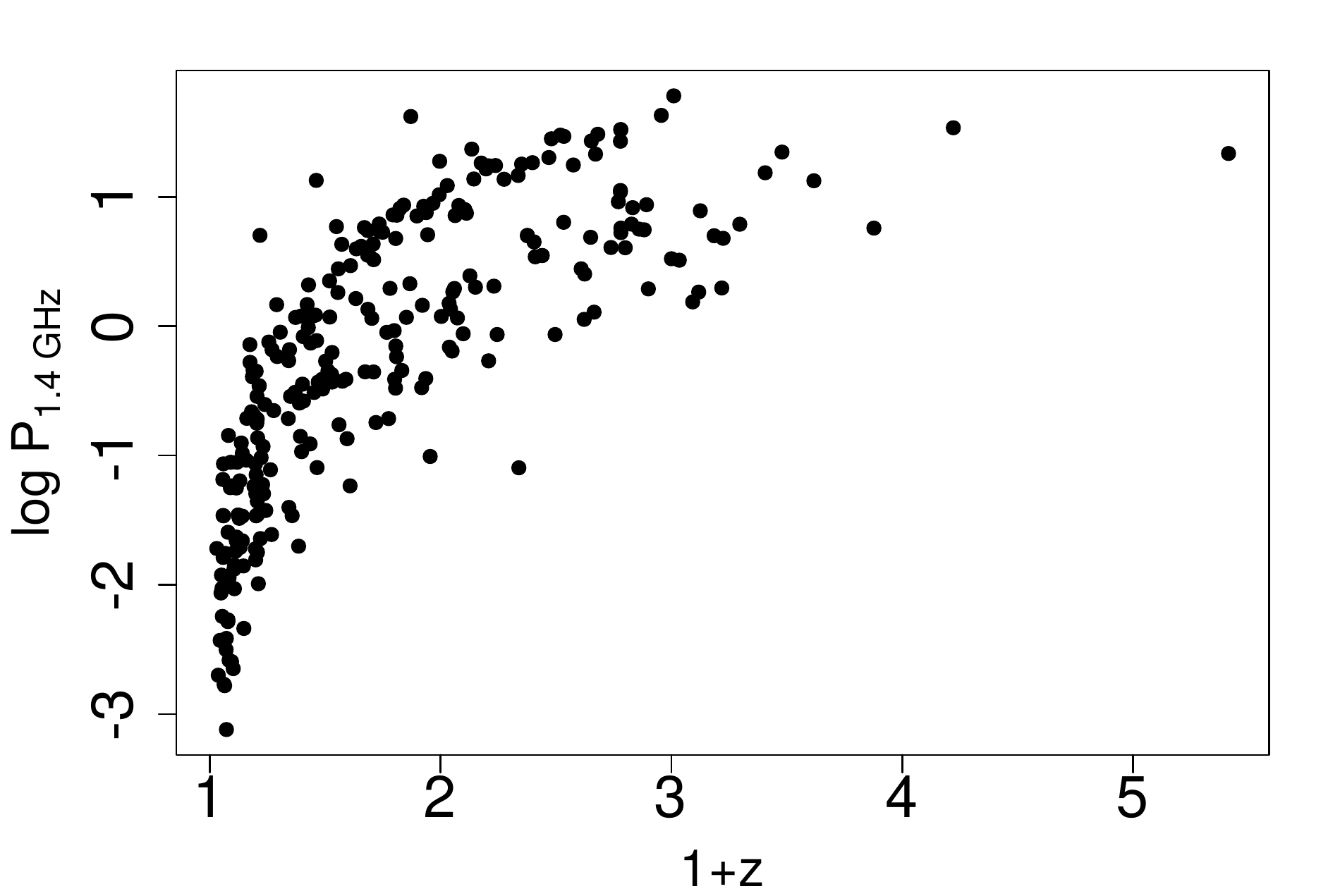}
    \caption{The distribution of the observed power spectral density at $1.4$\,GHz, in the units of $10^{26}$\,W\,Hz$^{-1}$\,sr$^{-1}$, with redshift, for the sample of 271 FR\,II sources analyzed in this paper.}
    \label{fig:P-z}
\end{figure}

Table\,\ref{tab:DYNAGE} presents the sample of 271 FR\,II sources analyzed in this paper, with their basic observed properties (power spectral density at $1.4$\,GHz, total linear size, aspect ratio of the radio lobes), as well as the corresponding DYNAGE model data (including the jet lifetime, total jet kinetic power, density of the gaseous atmosphere, injection spectral index, and the integrated spectral index between the emitted frequencies 0.4\,GHz and 5\,GHz). In Figure\,\ref{fig:P-z}, we plot the distribution of the $1.4$\,GHz power density with redshifts for the sources included in the sample. For the final revised sample, along with an in-depth discussion on the model fitting procedure, see Machalski et al. (2021, ApJS, accepted).

\section{Giant Radio Sources in the Sample} 
\label{ac}

\begin{figure}[!th]
\centering
\includegraphics[width=0.48\columnwidth]{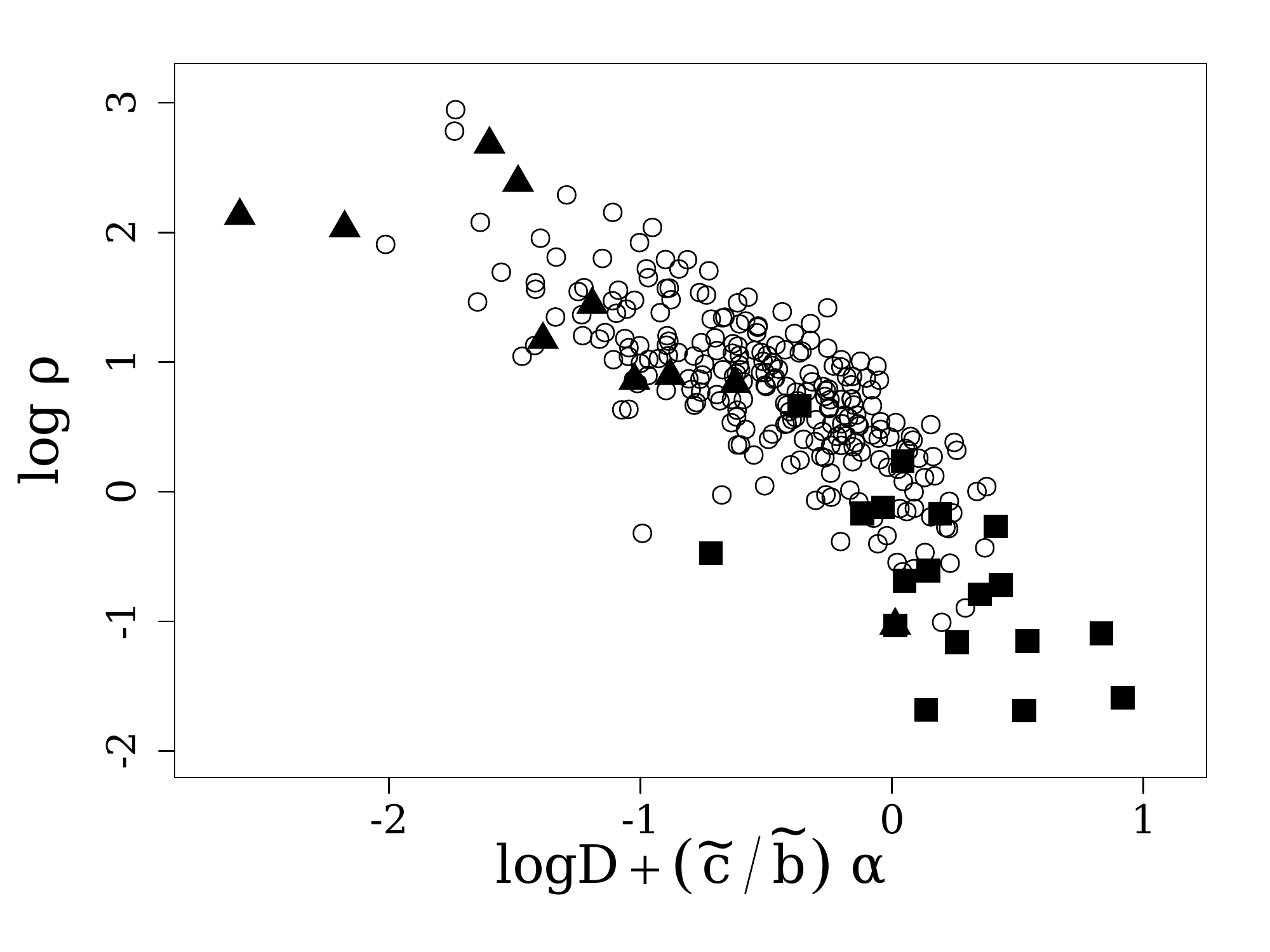}
\includegraphics[width=0.48\columnwidth]{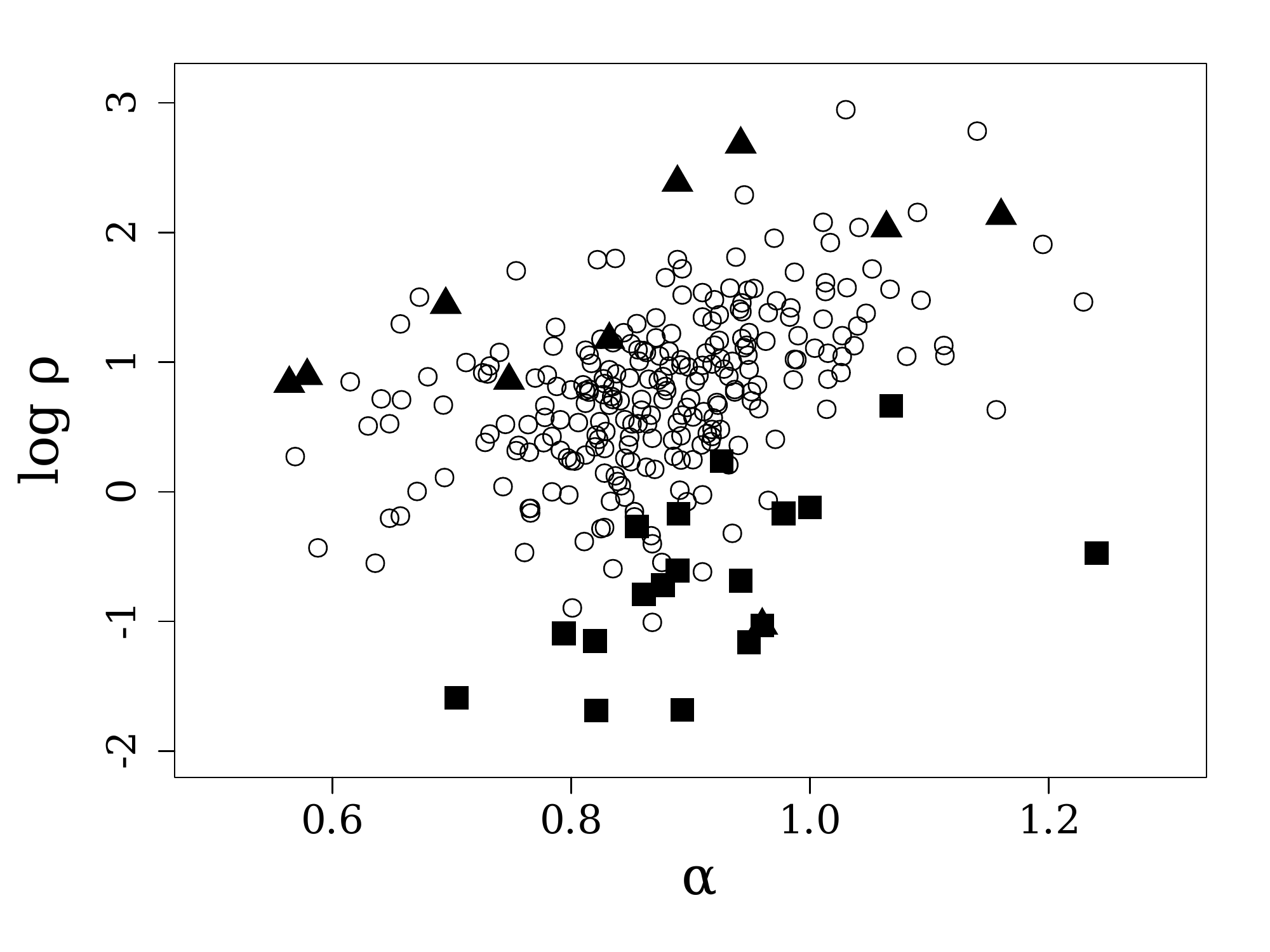}
\includegraphics[width=0.48\columnwidth]{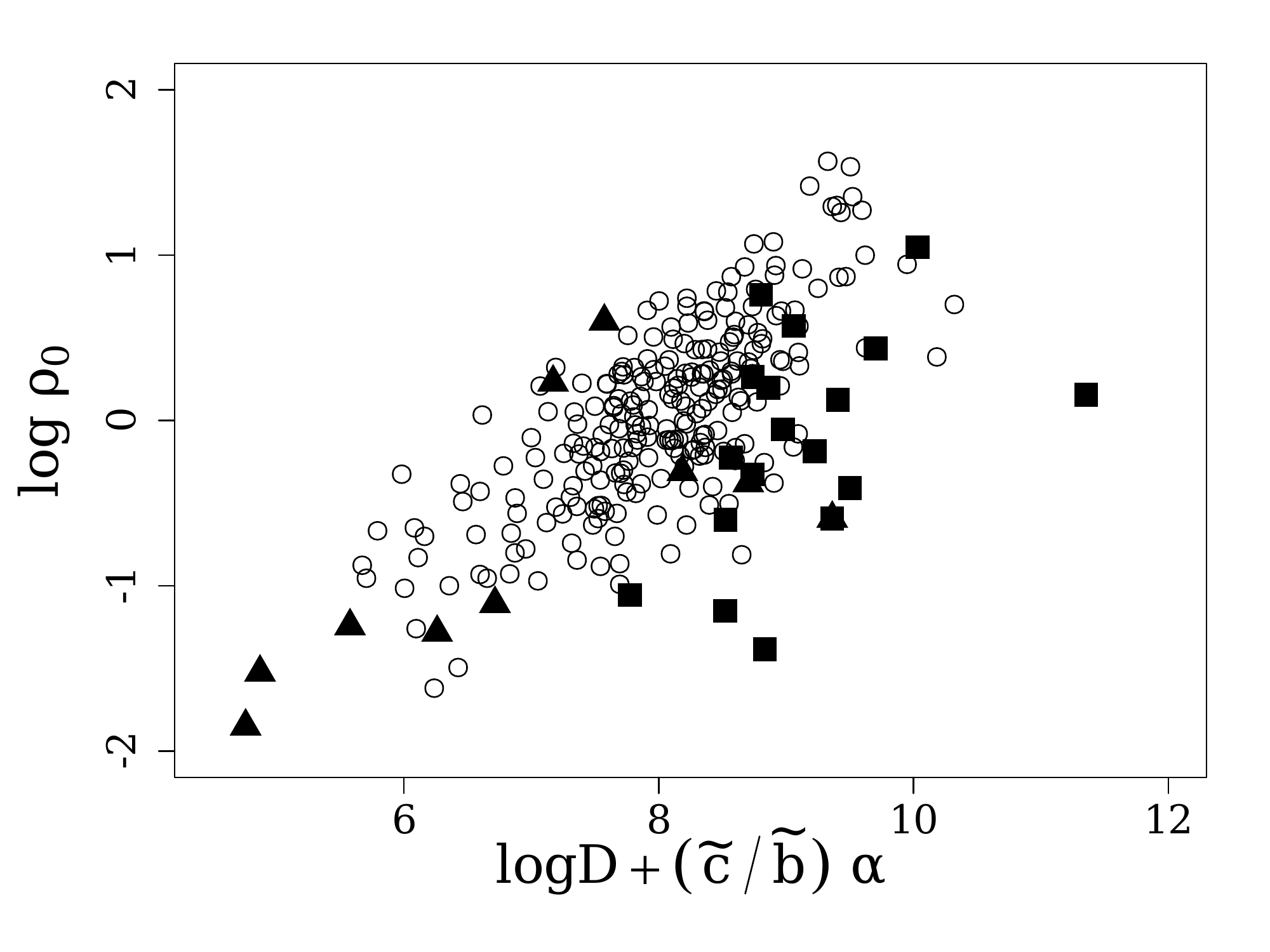}
\includegraphics[width=0.48\columnwidth]{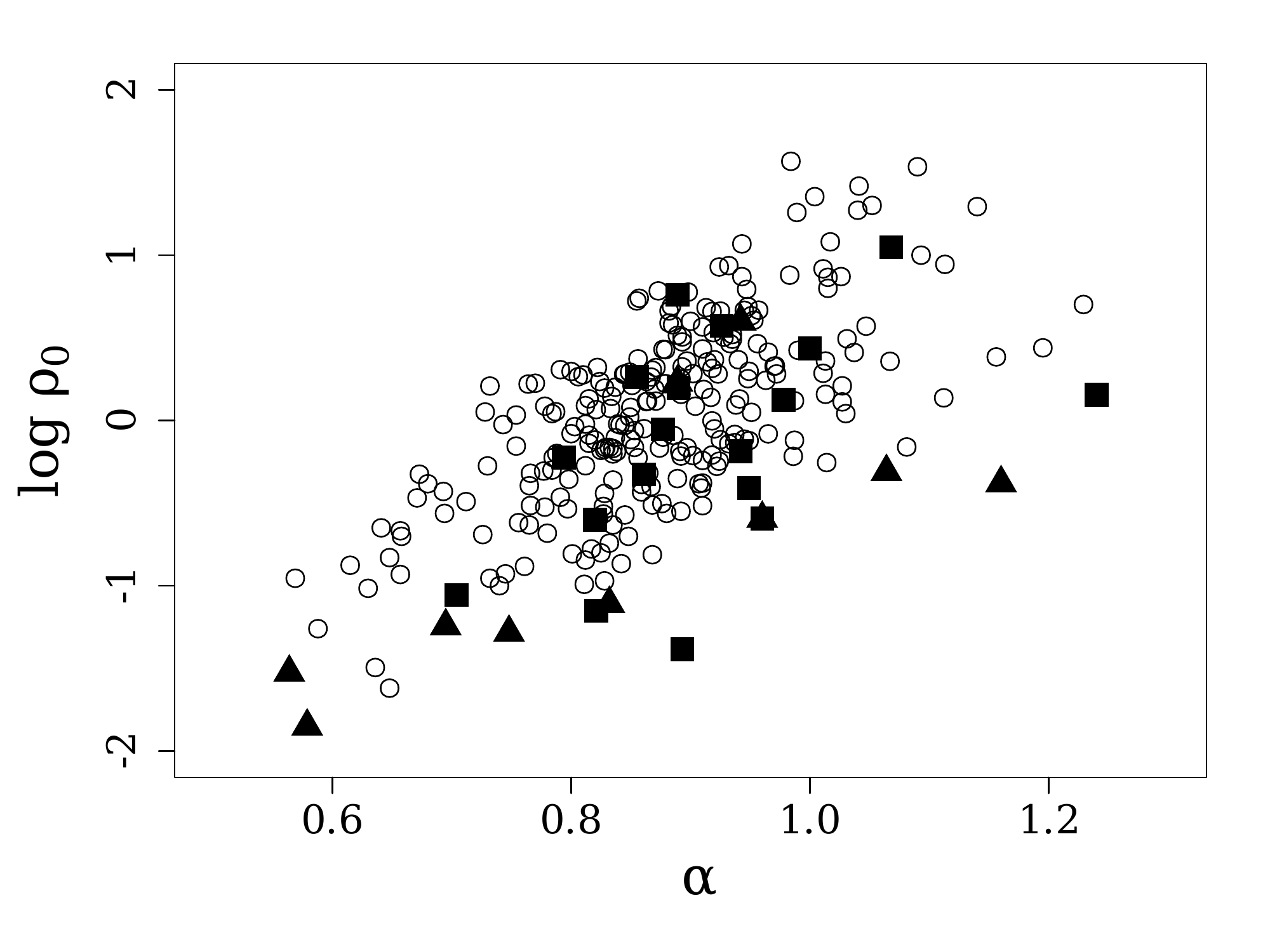}
\caption{Correlations between the model parameters for the entire sample of 271 sources (our full sample), with triangles representing the most compact sources with linear sizes $D<30$\,kpc, empty dots representing intermediate objects with $30$\,kpc\,$<D<1$\,Mpc, and squares denoting giant radio galaxies with $D>1$\,Mpc.}
 \label{fig:giants}
\end{figure}

In Figure\,\ref{fig:giants}, we present the studied correlations between the model parameters for the entire sample of 271 sources (our ''full sample"), with triangles representing the most compact sources with linear sizes $D<30$\,kpc, black dots representing intermediate objects with $30$\,kpc\,$<D<1$\,Mpc, and squares denoting giant radio galaxies with $D>1$\,Mpc. As follows, giant radio galaxies evolve in a substantially sparser ambient medium when compared to smaller sources, as in fact expected. Moreover, there is a gradual progression toward lower values of $\rho$ from the most compact objects to giant radio galaxies (see both upper panels in Figure\,\ref{fig:giants}); there is no such a trend when the central gas density $\rho_0$ is used instead (see the two lower panels in the figure), and this assures us that the DYNAGE algorithm captures correctly the dependencies between the model parameters. On the other hand, we also see that giant radio galaxies form an outlier population with respect to the correlations followed by all the other (smaller) sources.\\
Here, it is important to note that assuming the normality of the noise in a Bayesian regression analysis, the conflicting sources of information may contaminate the inference. This results in an undesirable effect-- the posterior will concentrate in an area in between the main population of data points and outliers with a scaling large enough to incorporate them all, what leads to erroneous predictions and also can enlarge the spread. This effect can be mitigated if a population of outlying sources can be dropped from the sample. Generally in order to minimize the influence of the outliers on our prediction, the modification of model is recommended, for instance replacing it with Student-t distribution that can efficiently identify and reject an outliers from the analysis, for more details see \citet{Martinez2017}.

Indeed, for the full sample, the residual density distributions (after the regression fit is made), deviate from normal. We quantify this finding by means of the Shapiro-Wilk and the Anderson-Darling normality tests, which are summarized in Table\,\ref{tab:normality}. As follows, in all the cases but $\rho_0(\alpha)$, for the full sample the distributions of the residuals vary from a Gaussian distribution on 95\% level of confidence (p-values $<0.05$). Meanwhile, for the $D<1$\,Mpc sub-sample, the p-values are all $>0.05$, meaning that the distributions of residuals are in this case consistent with normal.\\
We note also the double peaked residual distribution of the full sample, that seems to be smoothed when extracting the giant radio-sources that seems to confirm our hypothesis about the sub-populations of sources in our database.

\clearpage

\startlongtable
\begin{deluxetable}{ccccccccccc}
\tabletypesize{\footnotesize}
\tablecaption{The sample of the selected 271 FR\,II sources with the corresponding DYNAGE model data. \label{tab:DYNAGE}}
\tablewidth{0pt}
\tablehead{
\colhead{Name} & \colhead{$(1+z)$} & \colhead{$P_{\rm 1.4\,GHz}$} & \colhead{$D$} & \colhead{$A\!R$} & \colhead{$\tau$} & \colhead{$Q$} & \colhead{$\rho_0 a_0^{3/2}$} & \colhead{$\rho$} & \colhead{$\alpha_{\rm inj}$} & \colhead{$\alpha$} \\
\colhead{} & \colhead{} & \colhead{[$10^{26}$\,W\,Hz$^{-1}$\,sr$^{-1}$]} & \colhead{[kpc]} & \colhead{} & \colhead{[Myr]} & \colhead{[$10^{45}$\,erg\,s$^{-1}$]} & \colhead{[$10^8$\,g\,cm$^{-3/2}$]} & \colhead{[$10^{-28}$\,g\,cm$^{-3}$]} & \colhead{} & \colhead{} \\
\colhead{(1)} & \colhead{(2)} & \colhead{(3)} & \colhead{(4)} & \colhead{(5)} & \colhead{(6)} & \colhead{(7)} & \colhead{(8)} & \colhead{(9)} & \colhead{(10)} & \colhead{(11)}
}
\startdata
3C6.1&1.84&8.65&199&3&4&14.55&0.292&1.827&0.54&0.797\\
3C9&3.01&60.53&119&3&1.97&89.74&1.295&16&0.617&1.027\\
3C13&2.351&17.95&238&3&6.2&28.05&1.12&5.033&0.59&0.951\\
3C14&2.469&20.18&205&3&3.22&41.4&0.39&2.295&0.59&0.909\\
3C16&1.405&0.832&594&3.3&30&3.524&2.14&2.538&0.63&0.971\\
3C20&1.174&0.721&155&3&5.65&1.774&0.241&2.281&0.552&0.756\\
3C22&1.938&7.568&193&3&4.87&15.2&0.613&3.774&0.591&0.902\\
3C33&1.0595&0.0863&311&1.75&61&0.2&0.7&2.075&0.517&0.754\\
3C33.1&1.181&0.218&716&2.4&112&0.624&1.92&1.767&0.513&0.902\\
3C34&1.689&3.54&346&4.3&24&10.3&20&52.35&0.63&1.052\\
3C35&1.0673&0.0175&856&4.35&174&0.125&3.79&2.5&0.543&0.885\\
3C41&1.794&7.261&190&2.95&1.52&39.29&0.032&0.281&0.534&0.636\\
3C42&1.395&1.197&175&2.6&8.9&2.99&0.815&5.868&0.555&0.815\\
3C46&1.4373&0.741&946&4.3&128&2.723&22.6&12.81&0.579&1.004\\
3C55&1.7348&5.458&527&3.2&27.8&13.84&4.305&5.909&0.576&0.951\\
3C61.1&1.1878&0.457&578&2.9&58&0.979&1.3&1.545&0.559&0.863\\
3C65&2.176&18.28&154&2.3&3.4&32.03&0.532&4.814&0.515&0.812\\
3C68.1&2.238&17.5&569&3.9&9.8&44.05&0.685&0.838&0.582&0.897\\
3C68.2&2.575&17.66&203&3&6.2&72.15&5.03&29.13&0.72&1.229\\
3C79&1.2559&0.755&372&2.55&51&1.4&4.81&11.77&0.535&0.913\\
3C98&1.0306&0.0191&198&3&42&0.0635&1.615&9.343&0.521&0.732\\
3C109&1.3056&0.9&430&4.8&37&1.45&5.49&10.16&0.51&0.857\\
4C14.11&1.206&0.191&409&1.55&64&0.458&0.501&0.998&0.518&0.784\\
3C123&1.2177&5.047&144&3&2.85&11.35&0.255&2.536&0.575&0.823\\
3C132&1.214&0.345&82&3&4&1.14&0.596&13.29&0.57&0.785\\
3C169.1&1.633&1.637&343&2.8&24.7&4.081&3.19&8.839&0.575&0.928\\
3C172&1.5191&2.244&667&4.7&25.6&6.09&1.55&1.489&0.568&0.87\\
3C173.1&1.2921&0.583&282&1.95&44&1.02&2.71&9.445&0.562&0.91\\
3C184&1.994&10.4&39&3&0.62&16.2&0.305&22.29&0.55&0.91\\
3C184.1&1.1182&0.0887&409&1.7&86&0.211&0.832&1.735&0.521&0.8\\
3C186&2.063&7.178&13.8&3&0.145&38.44&0.418&134.7&0.785&1.16\\
3C190&2.197&16.48&58&3&0.74&30.43&0.282&10.44&0.596&0.892\\
3C191&2.956&42.76&53&3&1.16&43.27&2.13&90.32&0.555&0.97\\
3C192&1.0598&0.0342&238&2.45&55&0.0918&1.68&7.539&0.525&0.77\\
3C196&1.871&41.88&45&2&0.49&53.16&0.107&6.835&0.505&0.828\\
3C200&1.458&1.219&148&3&10.9&3.05&4.04&37&0.593&0.953\\
3C204&2.112&7.482&312&3.4&14.1&34.06&10&30.01&0.667&1.093\\
3C205&2.534&29.44&156&4.25&1.45&89.78&0.793&7.738&0.584&0.877\\
3C207&1.684&3.548&105&3&2.9&6.72&0.481&7.379&0.578&0.865\\
3C216&1.67&5.794&236&3&6.7&11.92&0.618&3.024&0.586&0.918\\
3C217&1.8975&7.128&94&4&1.7&14.18&0.679&15.21&0.525&0.943\\
3C219&1.1747&0.526&556&2.25&76&1.44&2.26&2.816&0.56&0.914\\
3C220.1&1.61&2.944&186&4&11&7.96&12.04&83.59&0.61&1.017\\
3C220.3&1.685&5.495&56&3.2&2.55&7.26&4.64&194.9&0.54&0.945\\
3C223&1.1368&0.125&732&2.9&97&0.503&1.59&1.327&0.537&0.837\\
4C73.08&1.0581&0.0163&1023&1.2&146&0.178&0.0412&0.0208&0.524&0.893\\
3C234&1.1849&0.406&359&2.4&61&0.973&4.89&11.31&0.586&0.948\\
3C239&2.781&33.11&102&1.35&2.9&80&0.691&11.09&0.673&1.081\\
3C244.1&1.428&2.089&314&2.7&24&4.36&3.31&10.14&0.572&0.935\\
3C245&2.029&12.25&78&3&0.96&24.66&0.181&4.653&0.57&0.832\\
3C247&1.7489&5.333&102&2.4&2.7&10.43&0.414&7.71&0.45&0.68\\
3C252&2.105&7.98&493&3&32&25.48&8.8&11.21&0.67&1.113\\
3C254&1.734&6.18&124&2&5.6&14.75&3.12&37.54&0.627&1.031\\
3C263&1.6563&4.15&401&2.75&14.3&12.45&1.83&3.897&0.557&0.867\\
3C263.1&1.824&8.147&55&1.8&2.25&7.71&0.567&23.19&0.544&0.924\\
3C265&1.8108&7.228&627&3.9&27.3&27.7&7.35&7.382&0.609&1.015\\
3C266&2.275&13.71&38&2.4&0.86&27.8&1.1&885.6&0.639&1.03\\
3C267&2.144&13.77&335&3.1&8.8&35.3&1.54&4.151&0.575&0.911\\
3C268.4&2.3978&18.41&121&3.6&3.4&20&2.329&30.29&0.51&0.92\\
3C270.1&2.519&30.06&134&1.7&2.35&54&0.199&2.294&0.568&0.848\\
3C272&1.944&5.105&490&3.25&19&13.6&1.78&2.702&0.557&0.892\\
3C274.1&1.422&1.472&856&3.7&64&4.58&4.59&3.024&0.565&0.925\\
3C275.1&1.557&2.78&181&4&7.5&4.24&1.721&11.88&0.53&0.863\\
3C280&1.996&18.88&116&1.45&2.6&33.1&0.143&1.921&0.56&0.812\\
3C280.1&2.6713&21.43&180&4&2.75&40.35&0.823&5.884&0.588&0.937\\
3C284&1.239&0.248&697&3.8&105&0.887&8.65&7.758&0.562&0.932\\
3C285&1.0794&0.0255&235&1.65&88&0.0465&1.352&6.195&0.525&0.815\\
3C289&1.967&8.933&86&1.25&3.6&12&0.274&6.037&0.571&0.88\\
3C292&1.71&3.273&960&4.2&84&8.66&18.13&10.44&0.528&0.989\\
3C293&1.0452&0.00372&247&3&61&0.0206&0.686&2.918&0.61&0.829\\
3C294&2.779&26.98&144&1.9&3.65&91.8&1.37&13.44&0.683&1.112\\
3C295&1.4614&13.43&34&3&0.31&21.99&0.1&11.89&0.48&0.74\\
3C300&1.27&0.658&436&2.4&52&1.31&2.29&4.472&0.522&0.897\\
3C303&1.1412&0.104&151&3&18.1&0.335&1.9&16.87&0.578&0.844\\
3C319&1.192&0.203&316&2.3&77&0.485&7.57&22.24&0.599&0.983\\
3C321&1.0961&0.0583&576&4&113&0.226&6.08&7.261&0.547&0.873\\
3C322&2.681&30.55&301&2.5&4&81.7&0.268&0.908&0.557&0.845\\
3C324&2.207&17.42&89&1.8&2.2&29.6&0.414&7.878&0.584&0.907\\
3C325&2.135&23.44&141&2.2&2.58&39.6&0.343&3.592&0.525&0.791\\
3C326&1.0895&0.0564&1804&3.5&196&0.481&0.883&0.19&0.499&0.877\\
3C330&1.549&5.9&396&5.45&6.6&25.7&1.1&2.672&0.543&0.784\\
3C334&1.555&1.824&429&2.3&49&4.576&6.303&11.71&0.61&1.015\\
3C336&1.927&8.472&184&3&5.5&18.12&1.439&9.516&0.57&0.892\\
3C337&1.635&3.972&306&2.7&11.3&9.98&0.761&2.219&0.553&0.82\\
3C341&1.448&1.202&487&4.9&30&4&5.98&9.186&0.565&0.898\\
3C349&1.205&0.287&312&3.25&29&0.807&1.98&6.114&0.535&0.8\\
3C351&1.3719&1.17&359&1.5&28&2.33&0.306&0.745&0.527&0.766\\
3C352&1.8057&4.775&104&2.5&4.38&10.73&1.912&29.75&0.612&0.972\\
3C356&2.079&8.59&653&3.95&42&21.7&18.7&19.02&0.57&1.04\\
3C381&1.1605&0.194&215&2.25&25&0.462&0.94&5.035&0.535&0.841\\
3C382&1.0578&0.0344&219&1.05&88&0.0585&0.403&2.018&0.512&0.765\\
3C388&1.0908&0.0887&85&1.2&21.6&0.138&0.627&14.12&0.56&0.835\\
3C390.3&1.0569&0.0653&262&1.4&85&0.117&1.398&5.428&0.537&0.834\\
3C401&1.2011&0.449&78&1.5&10.2&0.427&0.441&0.945&0.475&0.798\\
3C427.1&1.572&4.305&163&2.15&9.1&8.12&1.75&14.5&0.592&0.963\\
3C436&1.2145&0.347&397&4.3&50.4&1.11&11.7&24.46&0.58&0.943\\
3C437&2.48&28.18&312&3.95&2.26&130.3&0.233&0.743&0.562&0.765\\
3C438&1.29&1.469&103&2.67&15.4&1.99&2.29&35.04&0.592&1.013\\
3C441&1.708&4.325&262&2.6&10.2&9.64&0.933&3.598&0.556&0.845\\
3C452&1.0811&0.143&447&3.1&102&0.349&8.49&14.72&0.565&0.924\\
3C457&1.427&0.977&1056&3.75&81&4.08&3.74&1.726&0.558&0.926\\
3C469.1&2.336&14.66&666&4.7&13&63.32&2.33&2.278&0.584&0.94\\
3C470&2.653&27.1&218&4.2&2.14&95.9&0.595&3.351&0.576&0.856\\
6C0820+3642&2.86&5.636&196&3&7.54&16.22&2.66&15.99&0.606&0.99\\
6C0822+3417&1.406&0.264&119&1.5&17.5&0.451&0.687&8.738&0.557&0.832\\
6C0823+3758&1.2065&0.0179&250&2.35&179&0.055&26.19&109.4&0.614&1.041\\
6C0854+3956&1.528&0.424&1014&2.7&69&2.43&5.75&0.247&0.522&0.889\\
6C0857+3907&1.229&0.0596&100&3&11.2&0.199&1.13&18.58&0.554&0.787\\
6C0905+3955&2.882&5.572&68&3&1.45&58&2.75&80.94&0.784&1.195\\
6C0908+3736&1.1047&0.0132&77&1.15&19.8&0.034&0.148&3.347&0.514&0.648\\
6C0919+3806&2.65&4.875&94&3&2.87&11.42&1.35&25.64&0.593&0.941\\
6C0922+3640&1.1121&0.0182&407&2.7&111&0.062&1.84&3.417&0.511&0.806\\
6C0943+3958&2.037&1.503&92&2.65&5.3&3.06&1.79&35.88&0.587&0.948\\
6C0955+3844&2.405&4.487&208&3&6.92&11.8&1.22&7.089&0.568&0.904\\
6C1011+3632&2.043&1.358&443&3.5&19.4&5.68&1.38&2.437&0.57&0.917\\
6C1016+3637&2.892&8.71&206&3&2.23&32.67&0.117&0.65&0.49&0.657\\
6C1017+3712&2.053&1.845&104&3&6.35&3.44&3.1&0.478&0.562&0.935\\
6C1018+3729&1.806&0.703&603&3.2&62&3.14&7.42&8.305&0.61&1.026\\
6C1019+3924&1.921&1.452&79&3&4.25&2.98&2.11&52.46&0.55&0.893\\
6C1031+3405&2.832&8.26&381&3&4.66&34.73&0.131&0.34&0.515&0.761\\
6C1042+3912&2.77&9.183&79&3&0.7&30.57&0.0965&3.217&0.48&0.63\\
6C1045+3403&2.827&6.166&154&3&3.85&13.32&0.676&6.485&0.53&0.835\\
6C1100+3505&2.44&3.524&119&3.8&2.67&11.45&0.892&12.28&0.574&0.861\\
6C1108+3956&1.59&0.389&117&3&7&1.19&0.95&12.36&0.564&0.812\\
6C1113+3458&3.406&15.38&132&3.5&1.47&40.95&0.299&3.745&0.53&0.778\\
6C1129+3710&2.06&1.959&127&2.65&6.3&3.88&1.2&13.87&0.547&0.85\\
6C1130+3456&1.512&0.451&524&3.8&34&1.75&1.57&2.156&0.545&0.828\\
6C1134+3656&3.125&7.816&155&3.2&2.55&30.79&0.528&4.881&0.61&0.922\\
6C1141+3525&2.781&5.741&116&2.3&3.6&11.89&0.89&13.51&0.589&0.92\\
6C1148+3638&1.1412&0.0219&131&1.1&107&0.0262&2.92&37.29&0.546&0.933\\
6C1158+3433&1.53&0.37&267&4&14.6&1.42&1.22&4.612&0.546&0.778\\
6C1204+3708&2.779&10.94&502&4&7.65&48.08&0.65&1.028&0.556&0.891\\
6C1204+3519&2.376&5.047&153&3&6.55&10.1&2.59&24.07&0.573&0.965\\
6C1220+3723&1.489&0.327&225&3&17.6&0.99&1.23&6.022&0.553&0.812\\
6C1230+3459&2.533&6.368&106&3&2.4&14.98&0.681&11.15&0.573&0.874\\
6C1232+3942&4.221&34.28&85&3&1.6&51.3&19.68&605.5&0.62&1.14\\
6C1256+3648&2.127&2.455&131&3.4&3.2&8.78&0.632&6.497&0.552&0.788\\
6C1257+3633&2.003&1.191&343&3.6&28.1&4.26&8.27&21.48&0.609&1.011\\
6C1301+3812&1.47&0.37&214&3.4&17.7&1.08&2.36&12.44&0.564&0.856\\
5C6.5&2.038&0.689&186&3&13.9&0.982&1.05&7.551&0.503&0.849\\
5C6.17&2.05&0.643&396&3&26.3&1.53&0.723&1.614&0.525&0.932\\
5C6.19&1.799&0.925&68&3&2.02&1.87&0.204&8.276&0.49&0.726\\
5C6.24&2.073&1.159&11.1&3&0.14&2.7&0.057&27.79&0.53&0.695\\
5C6.29&1.72&0.18&92.4&3&4.7&0.634&0.302&5.616&0.61&0.827\\
5C6.33&2.496&0.863&162&3&9.2&1.98&0.994&9.663&0.575&0.918\\
5C6.34&3.118&1.837&94&3.9&1.1&7.4&0.118&3.31&0.566&0.745\\
5C6.63&1.465&0.0807&371&3&43&0.288&0.812&1.876&0.59&0.886\\
5C6.75&1.775&0.193&126&3.2&12.2&0.382&1.31&15.39&0.568&0.871\\
5C6.78&1.263&0.0776&1460&6&106&0.798&1.83&0.54&0.51&0.855\\
5C6.83&2.8&4.046&134&3&3.7&6.48&0.41&5.157&0.535&0.859\\
5C6.95&3.877&5.741&160&3&2.3&17.56&0.136&1.112&0.562&0.842\\
5C6.160&2.624&2.535&76&3&1.3&5.48&0.111&2.789&0.546&0.732\\
5C6.214&1.595&0.135&216&3&26.5&0.588&2.58&13.39&0.675&1.037\\
5C6.217&2.41&3.443&103&3&0.79&15.32&0.024&0.625&0.53&0.648\\
5C6.233&1.56&0.173&48.3&3&4.3&0.349&1.24&64.6&0.593&0.938\\
5C6.239&1.805&0.332&1073&6&58&1.3&1.33&0.679&0.506&0.978\\
5C6.242&2.9&1.945&42.6&3&0.86&3.94&0.167&9.777&0.59&0.817\\
5C6.251&2.665&1.285&71.4&3&2.03&2.87&0.272&7.442&0.579&0.827\\
5C6.264&1.832&0.455&57.1&3&1.93&1.17&0.208&7.965&0.575&0.78\\
5C6.267&1.357&0.0343&23.7&3&3.9&0.053&1.7&242.9&0.575&0.889\\
5C6.286&2.339&0.0804&140&3&6.6&1.97&0.607&7.3&0.571&0.986\\
5C6.287&3.296&6.152&104&3&3.9&12.4&2.28&36.5&0.625&1.067\\
5C6.292&2.245&0.863&41.4&3&2.45&1.72&1.93&119.8&0.633&1.011\\
7C0221+3417&1.852&1.172&141&3&5&2.82&0.369&4.259&0.567&0.859\\
5C7.7&1.435&0.123&15.6&3&0.29&0.513&0.0299&6.834&0.493&0.564\\
5C7.8&1.673&0.444&320&3&19&1.49&0.609&1.756&0.61&0.892\\
5C7.9&1.233&0.0506&530&4&82&0.159&1.96&2.645&0.542&0.849\\
5C7.10&3.185&5.012&181&3&6.65&12.4&1.62&11.06&0.632&1.027\\
5C7.17&1.936&0.394&692&4&25&2.29&0.314&0.285&0.565&0.876\\
5C7.23&2.098&0.875&264&3&11&2.33&0.362&1.392&0.561&0.828\\
5C7.57&2.622&1.13&635&5&18.8&5.89&0.83&0.858&0.572&0.965\\
5C7.70&3.617&13.34&19.5&3&0.14&29.7&0.0778&14.91&0.602&0.832\\
5C7.78&2.151&2.005&188&3&15.2&2.77&3.72&23.81&0.572&1.047\\
5C7.79&1.608&0.0583&1864&6&102&1.65&1.43&0.336&0.52&1.24\\
5C7.82&1.918&0.335&358&3.8&30&0.92&1.91&4.655&0.568&0.923\\
5C7.85&1.955&0.0984&255&3&5.12&5.76&0.102&0.413&0.63&0.811\\
5C7.95&2.208&0.54&543&5&34&2.77&3.98&5.187&0.63&0.9\\
5C7.125&1.801&0.389&120&3&6.5&0.735&0.372&4.665&0.48&0.693\\
5C7.145&1.343&0.0397&104&2.4&17.9&0.08&0.727&11.31&0.56&0.815\\
5C7.170&1.268&0.0245&137&3&39.2&0.046&3.2&32.95&0.566&0.893\\
5C7.194&2.738&4.064&24.2&3&0.21&11.26&0.0521&7.222&0.576&0.748\\
5C7.195&3.034&3.243&31.1&3&0.49&6.46&0.158&15.03&0.587&0.825\\
5C7.205&1.71&0.443&120&3&7.35&0.89&0.648&8.133&0.569&0.838\\
5C7.208&3&3.327&165&3&6.05&5.49&0.735&6.145&0.554&0.937\\
5C7.223&3.092&1.542&42&3&1.4&3.81&0.759&49.39&0.6&0.987\\
5C7.245&2.609&2.78&112&3&3.85&5.72&0.765&10.66&0.597&0.925\\
5C7.269&3.218&1.977&69.5&3&2.9&4.73&1.44&40.99&0.636&1.013\\
5C7.271&3.224&4.786&9.6&3&0.041&16.1&0.0141&7.845&0.48&0.579\\
B0136+396&1.2107&0.0102&1174&3&273&0.757&11.19&4.591&0.636&1.068\\
B0211+393&1.198&0.019&386&2.5&167&0.0621&6.212&13.5&0.56&0.947\\
B0246+393&1.204&0.178&542&2.3&109&0.418&2.99&3.908&0.544&0.893\\
B0246+428A&1.159&0.092&404&2.7&98&0.217&4.6&9.338&0.541&0.882\\
J0654+7319&1.1145&0.022&1527&3.7&85&0.282&0.088&0.0257&0.477&0.704\\
B0724+506&1.35&0.287&758&2.4&69&1.099&0.668&0.53&0.52&0.828\\
B0811+388&1.132&0.0195&91.4&1.8&43.3&0.032&3.27&61.82&0.559&0.889\\
B0818+472A&1.1303&0.0637&34.9&1.5&7&0.101&0.787&63.09&0.562&0.837\\
B0823+379&1.2065&0.0348&250&2.35&151&0.121&34.28&142.9&0.652&1.09\\
B0828+324&1.0507&0.0119&315&2.45&52&0.0585&0.34&1.006&0.501&0.671\\
B0834+450A&1.2075&0.137&426&2&69&0.351&0.918&1.724&0.522&0.803\\
B0836+290&1.0791&0.00536&637&4.8&126&0.0492&2.03&2.088&0.501&0.791\\
B0844+319&1.0675&0.0114&367&1.95&139&0.0349&1.164&2.737&0.514&0.821\\
B0910+353&1.2197&0.0228&411&3.9&64.5&0.108&1.66&3.296&0.51&0.764\\
B0913+385&1.0711&0.00315&50&1.55&30.6&0.0056&1.079&50.66&0.525&0.754\\
J0926+610&1.243&0.0376&812&3.8&99&0.244&1.18&0.844&0.509&0.833\\
J0949+7314&1.0581&0.0172&982&1.5&200&0.0915&0.154&0.0983&0.512&0.868\\
B0955+320&1.201&0.071&549&2.5&115&0.211&2.016&2.586&0.53&0.868\\
B1005+282&1.1478&0.0046&638&3.35&127&0.0407&0.687&0.703&0.499&0.853\\
B1012+488&1.385&0.0199&610&3.05&92&0.721&3.4&3.714&0.549&0.919\\
B1013+410&1.1279&0.0327&287&2.5&102&0.0766&4.91&16.62&0.545&0.884\\
B1017+487&1.053&0.00938&390&3.15&88&0.0423&1.125&2.404&0.501&0.728\\
B1024+485&1.2311&0.1175&303&2.55&65&0.268&3.89&12.13&0.548&0.882\\
B1102+304&1.072&0.00385&231&1.7&78&0.014&0.274&1.287&0.496&0.694\\
B1104+365&1.3924&0.258&135&2.35&17.4&0.542&2.09&21.97&0.561&0.871\\
B1107+379&1.3456&0.659&405&3&32&1.742&1.72&3.475&0.54&0.824\\
B1113+295&1.0489&0.00865&74.4&1.3&21.7&0.021&0.199&5.121&0.518&0.658\\
B1130+339&1.2227&0.0964&105&2.2&11.6&0.242&0.531&8.098&0.532&0.73\\
B1141+374&1.1154&0.0561&587&3&80&0.224&0.944&1.095&0.501&0.743\\
B1141+466&1.1159&0.0234&27&1.95&9.5&0.0378&3.98&479.4&0.59&0.942\\
B1151+384B&1.1982&0.0865&227&1.25&53&0.164&0.494&2.39&0.523&0.777\\
B1204+341&1.0788&0.0052&80&1.6&20.5&0.0185&0.224&5.207&0.513&0.641\\
B1209+745&1.107&0.0142&795&2.9&152&0.0953&0.869&0.64&0.512&0.853\\
J1211+7419&1.107&0.00933&936&2.4&223&0.052&0.418&0.241&0.518&0.91\\
B1216+507&1.1995&0.0507&671&3.3&108&0.229&1.914&1.816&0.522&0.845\\
B1232+414A&1.1918&0.0579&279&2.5&58&0.144&1.888&6.686&0.529&0.81\\
B1247+336&1.463&0.7745&510&2.5&29&2.393&0.479&0.686&0.525&0.766\\
B1309+412A&1.1106&0.0137&742&1.85&121&0.0993&0.156&0.127&0.487&0.801\\
B1323+370&1.0545&0.0057&32.3&1.6&7.3&0.0158&0.215&19.71&0.524&0.657\\
B1347+285&1.0724&0.00076&68&1.45&69.2&0.0035&2.094&61.68&0.535&0.822\\
B1358+305&1.206&0.0443&2055&3.1&175&0.65&0.39&0.0691&0.527&0.949\\
B1405+517&1.3404&0.1932&675&2.8&118&0.677&4.65&4.373&0.569&0.957\\
B1422+268&1.037&0.002&99&2&18.9&0.0125&0.111&1.868&0.501&0.569\\
B1441+262&1.0621&0.0017&264&2.4&263&0.0056&7.415&28.65&0.551&0.943\\
B1452+502A&1.3883&0.2547&402&2.45&51.6&0.675&1.63&3.337&0.539&0.851\\
B1455+28C&1.1411&0.034&557&2.15&129&0.118&0.953&1.199&0.515&0.839\\
B1457+292&1.146&0.014&183&1.3&28&0.0631&0.055&0.369&0.495&0.588\\
B1519+512&1.277&0.2228&1210&2.9&248&0.714&0.65&0.206&0.51&0.942\\
B1528+291&1.0839&0.00261&348&2.1&188&0.0105&1.31&3.337&0.505&0.864\\
B1529+357&1.342&0.542&72.5&2.65&5.5&1.127&1.671&44.85&0.572&0.879\\
B1539+343&1.4018&0.3565&359&2.6&46.5&0.867&2.68&6.488&0.548&0.879\\
B1543+845&1.201&0.0342&1509&3.85&97&0.472&0.251&0.0705&0.493&0.82\\
B1602+324&1.452&0.3076&380&2.3&36&0.916&0.766&1.706&0.546&0.85\\
B1609+312&1.0944&0.00255&46&1.25&15.8&0.0075&0.133&7.051&0.507&0.615\\
B1613+275&1.0647&0.00166&39.3&1.75&16.8&0.0045&0.473&31.71&0.519&0.673\\
B1643+274&1.1018&0.00225&268&3.4&151&0.0086&5.285&19.82&0.519&0.855\\
B2303+391A&1.2061&0.0178&595&1.75&112&0.1&0.233&0.255&0.495&0.835\\
4C23.56&3.479&22.23&443&4.3&8.2&93.5&2.423&4.288&0.668&1.156\\
6C0140+326&5.413&21.68&17.7&2.7&0.24&34.5&0.489&108.4&0.65&1.064\\
J0042-0613&1.123&0.0347&760&2.7&113&0.183&0.659&0.519&0.529&0.825\\
J0202-0939&1.767&0.8974&1140&5.5&38.8&7.248&1.576&0.675&0.546&0.89\\
B0709+370&1.487&0.389&420&5&32&0.914&2.689&5.15&0.565&0.877\\
B0719+362&1.199&0.0156&66&3&7.7&0.0475&0.323&9.93&0.548&0.712\\
B0740+393&1.702&1.151&82&2.1&5.9&1.73&0.755&16.9&0.584&0.949\\
B0803+488&1.37&0.3083&546&2.9&72&0.665&2.065&2.67&0.566&0.918\\
B0809+360&1.6852&1.352&241&3&16.8&2.61&1.984&8.75&0.593&0.949\\
J0947-1338&1.08&0.0116&2451&11&99&0.512&0.595&0.081&0.517&0.794\\
B1014+397&1.53&0.6266&815&8&74&3.226&36.99&26.23&0.58&0.984\\
B1020+486&2.231&2.042&374&5.2&16&5.08&2.916&6.65&0.583&0.956\\
B1049+488&1.8093&0.5807&125&2.3&8.3&1.041&0.436&5.15&0.563&0.835\\
B1100+350&2.44&3.524&167&3&5.15&7.102&0.445&3.407&0.564&0.889\\
B1105+392&1.781&1.963&547&4.6&12&10.26&0.308&0.397&0.59&0.868\\
J1130+0630&1.398&0.1072&1476&3.2&60&0.965&0.0705&0.0205&0.512&0.821\\
B1141+354&2.781&5.297&99&2.6&3.3&8.99&0.771&12.92&0.596&0.945\\
B1148+477&1.867&2.133&165&1.2&19.5&2.636&0.556&4.329&0.573&1.014\\
B1204+353&2.376&5.012&162&3.8&5&10.52&1.316&10.5&0.579&0.987\\
B1204+371&2.779&11.22&464&4.8&7.1&33.44&0.573&0.947&0.567&0.91\\
J1334-1009&1.0838&0.0113&1322&3.3&144&0.255&0.472&0.162&0.529&0.861\\
B1339+472&1.502&0.537&146&3.5&11.5&1.495&3.676&34.39&0.581&0.91\\
J1434+0930&1.575&0.3758&1523&3.6&126&2.881&2.734&0.759&0.565&1\\
B1608+330&2.78&33.19&590&3.4&6.35&137.7&0.398&0.458&0.568&0.867\\
B1641+375&1.52&1.178&236&4.2&13.8&3.624&4.563&20.77&0.584&0.918\\
J2320-1320&1.3927&0.1406&1268&2.8&58&2.841&0.255&0.0932&0.65&0.96\\
\hline
\enddata
\tablecomments{Col.\,(1): name of the source; Col.\,(2): spectroscopic redshift; Col.\,(3): observed power spectral density at $1.4$\,GHz; Col.\,(4): total linear size of the radio structure; Col.\,(5): aspect ratio of the radio lobes; Col.\,(6): jet lifetime; Col.\,(7): total jet kinetic power; Col.\,(8): central density of the gaseous atmosphere; Col.\,(9): gas density at distances $D/2$ from the center of the host galaxy; Col.\,(10): injection spectral index; Col.\,(10): integrated spectral index of the lobes' radio emission, derived from the model spectra between the emitted frequencies 0.4\,GHz and 5\,GHz.}
\end{deluxetable}

\begin{deluxetable}{ccccc}[!h]
\tabletypesize{\footnotesize}
\tablecaption{Normality tests for the residual distributions. \label{tab:normality} }
\tablewidth{0pt}
\tablehead{
\colhead{Dataset} & \colhead{Model} & \colhead{SW statistics} & \colhead{AD statistics} & \colhead{Residual distribution} \\
\colhead{(1)} & \colhead{(2)} & \colhead{(3)} & \colhead{(4)} & \colhead{(5)}
}
\startdata
\\
full sample & $\log\!\rho \sim \mathcal{N}\!\left(a + b \, \log\!D + c \, \alpha,\sigma\right)$ & $W=0.98$ & $A=0.78$ & \includegraphics[width=2.5cm]{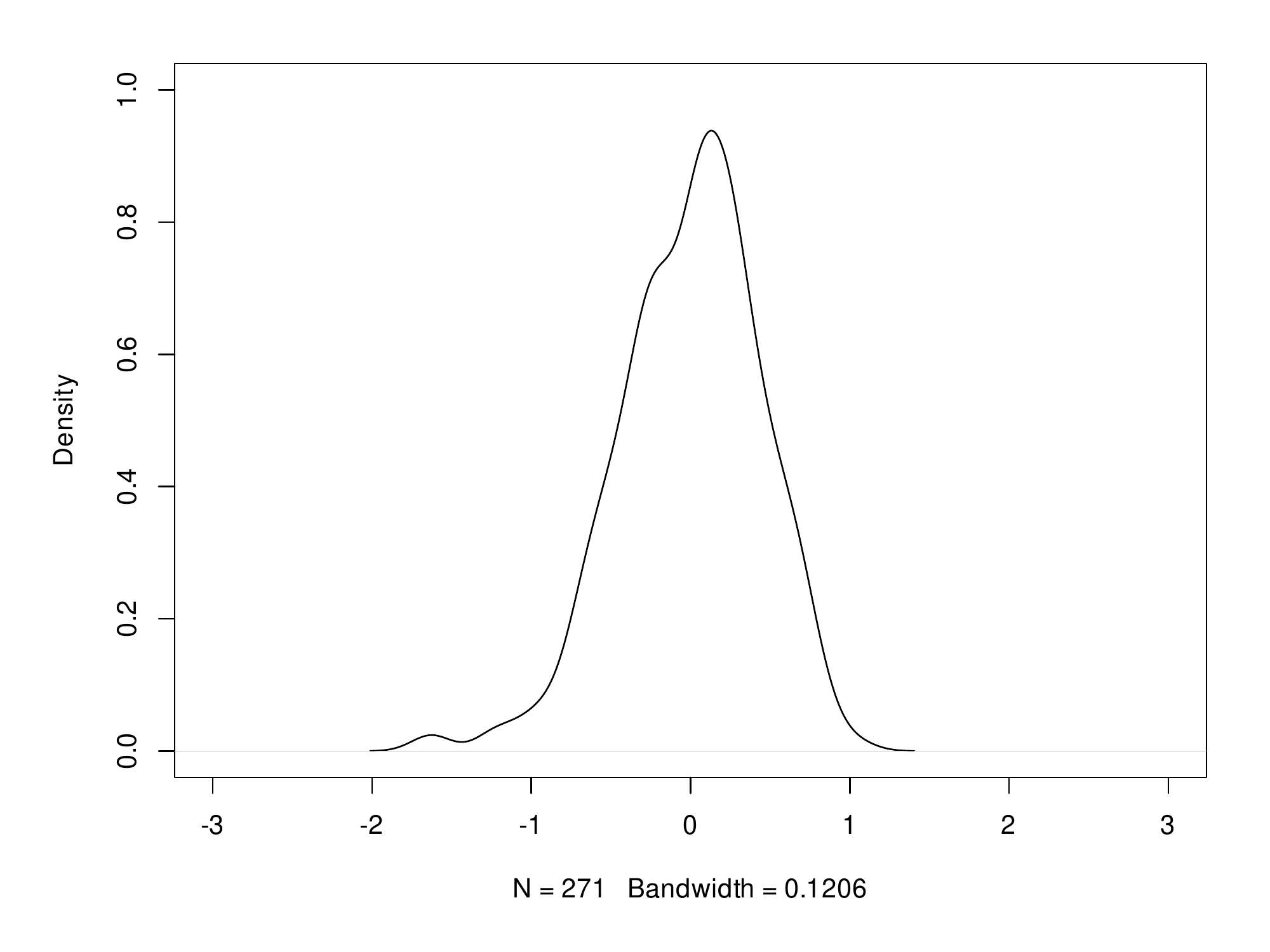}\\
 & & $p=0.0014$ & $p=0.04$ & \\
 \hline
$D<1$\,Mpc &  $\log\!\rho \sim \mathcal{N}\!\left(a + b \, \log\!D + c \, \alpha,\sigma\right)$ & $W=0.99$ & $A=0.27$ & \includegraphics[width=2.5cm]{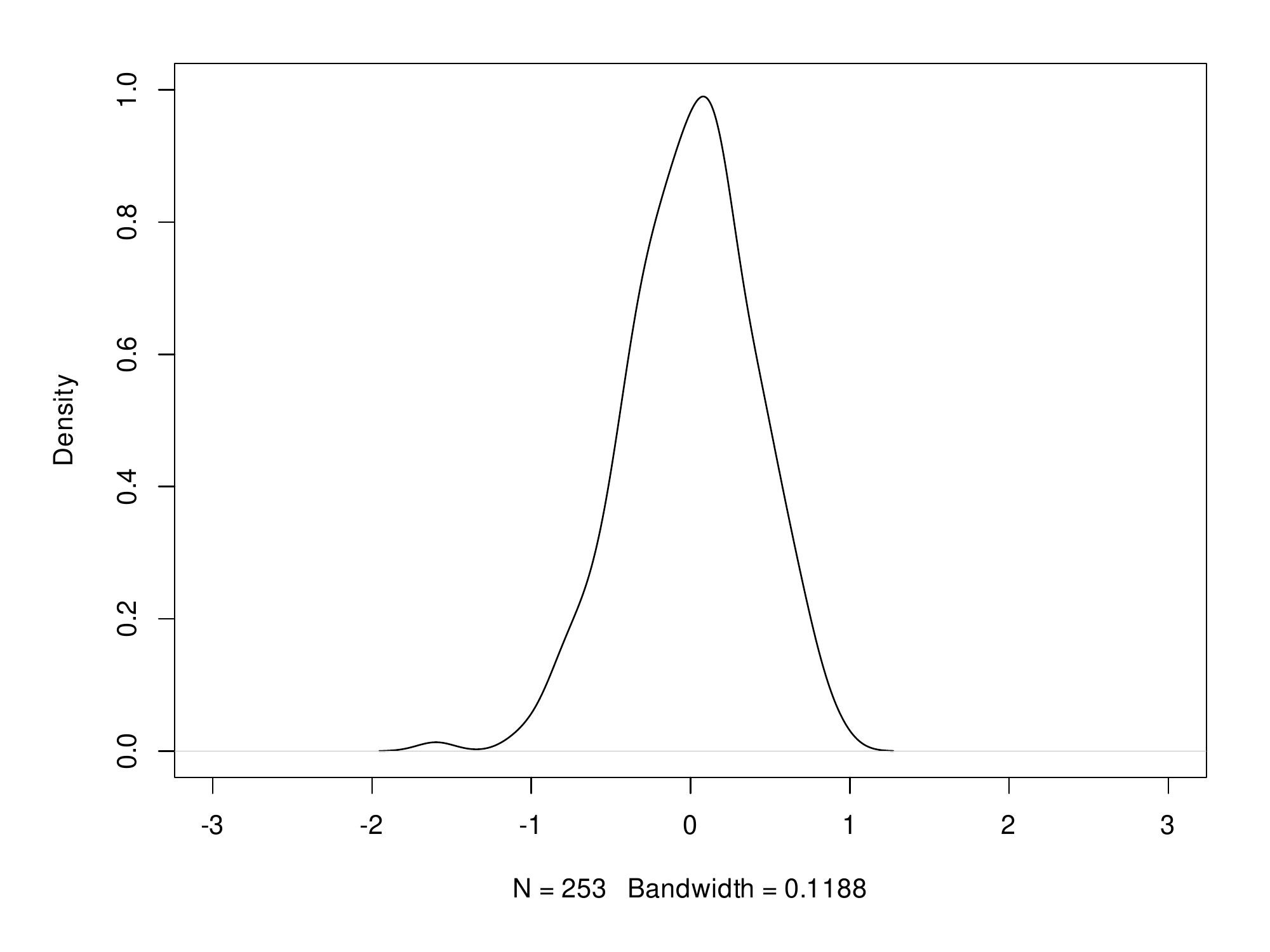}\\
 & & $p=0.11 $ & $p=0.67$ & \\
 \hline
full sample &    $\log\!\rho \sim \mathcal{N}\!\left(a + b \, \alpha;\,\sigma\right)$ & $W=0.97$ & $A=2.45$ & \includegraphics[width=2.5cm]{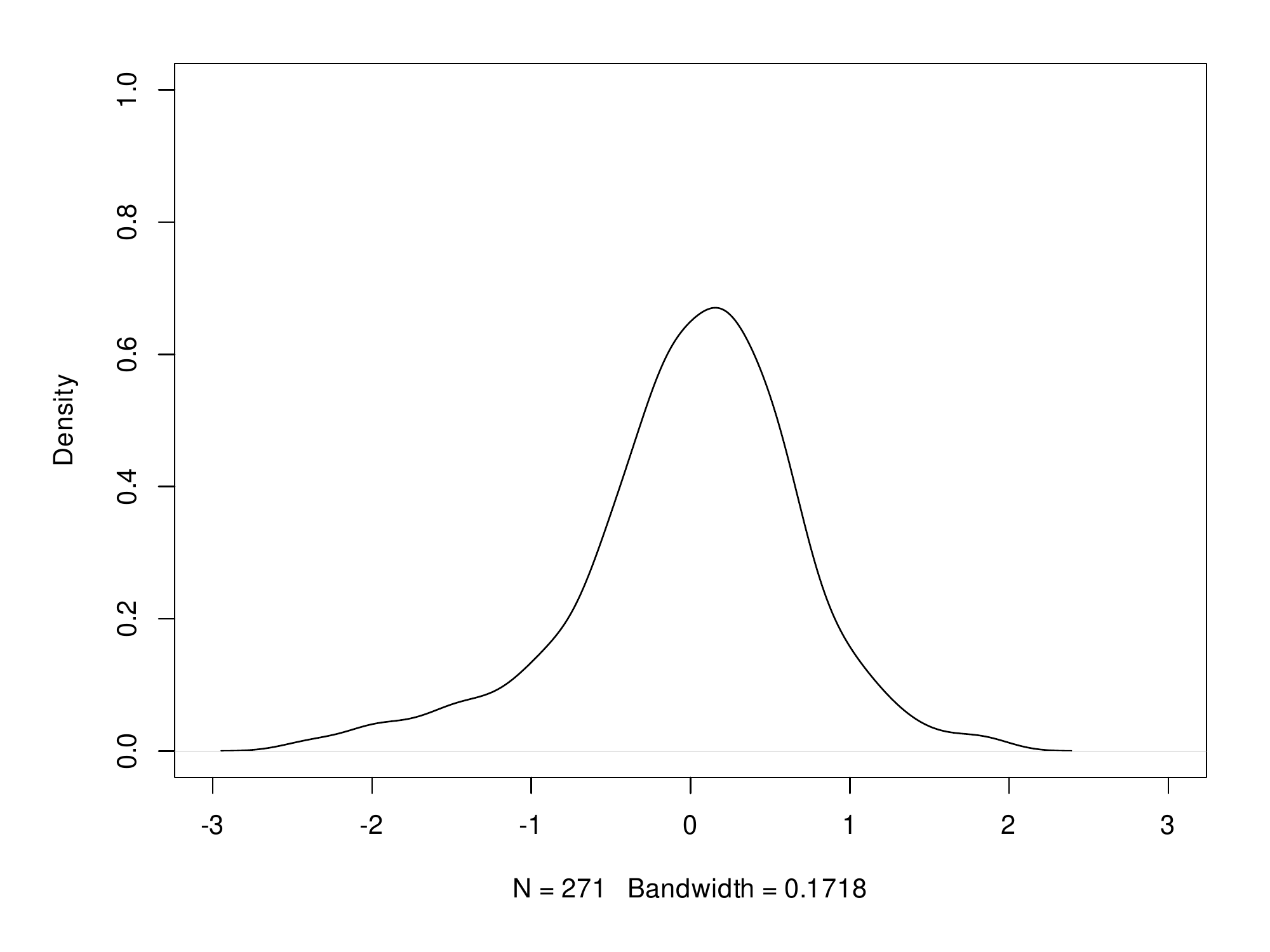}\\
 & & $p=1.4e-5$ & $p=3.3e-6$ & \\
 \hline
$D<1$\,Mpc &  $\log\!\rho \sim \mathcal{N}\!\left(a + b \, \alpha;\,\sigma\right)$ & $W=0.99$ & $A=0.44$ & \includegraphics[width=2.5cm]{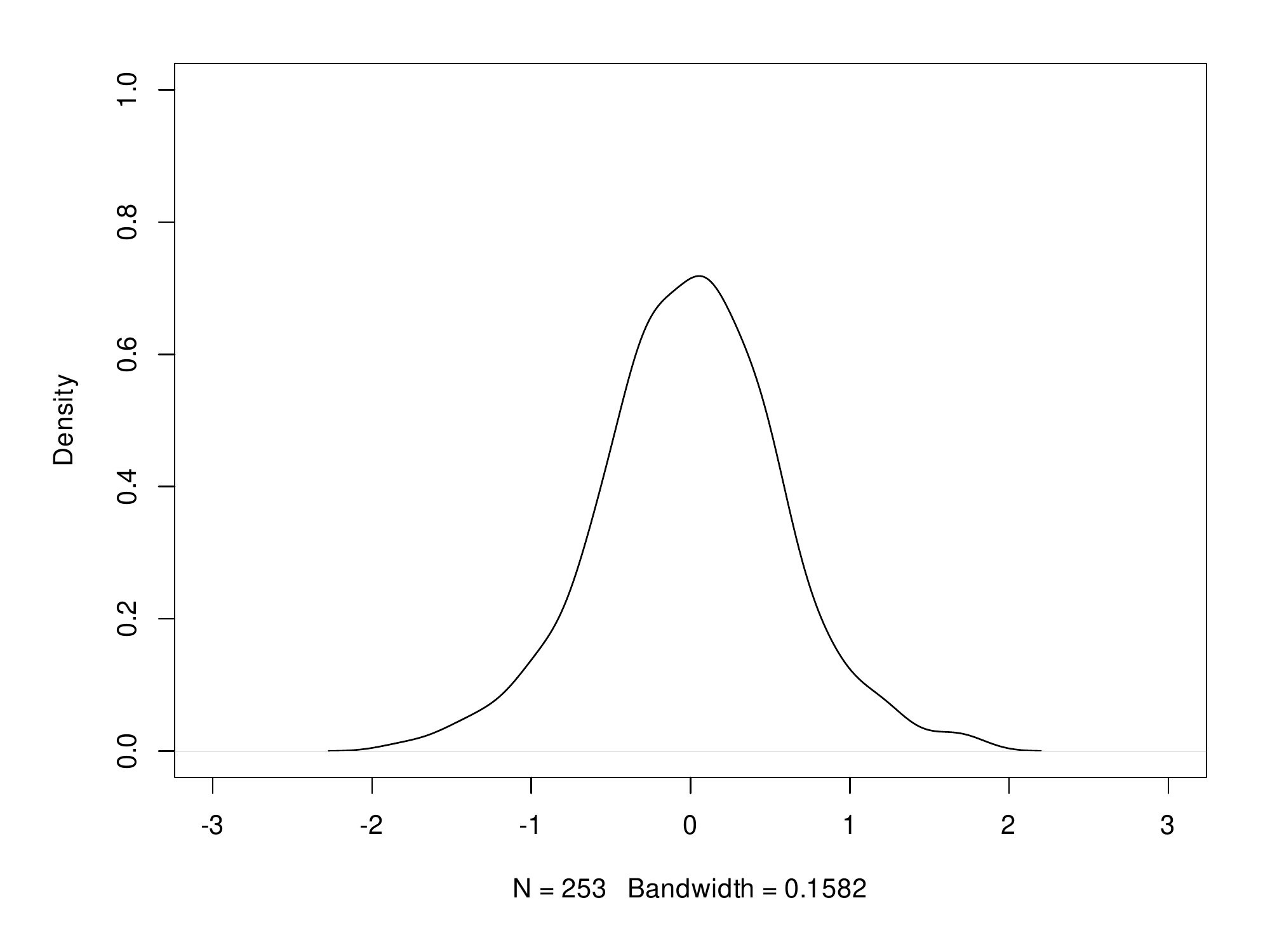}\\
 & & $p=0.37 $ & $p=0.29$ &\\
 \hline
full sample &  $\log \rho_0 a_0^{3/2} \sim \mathcal{N}\!\left(a + b \, \log\!D + c \, \alpha,\sigma\right)$ & $W=0.99$ & $A=0.67$ & \includegraphics[width=2.5cm]{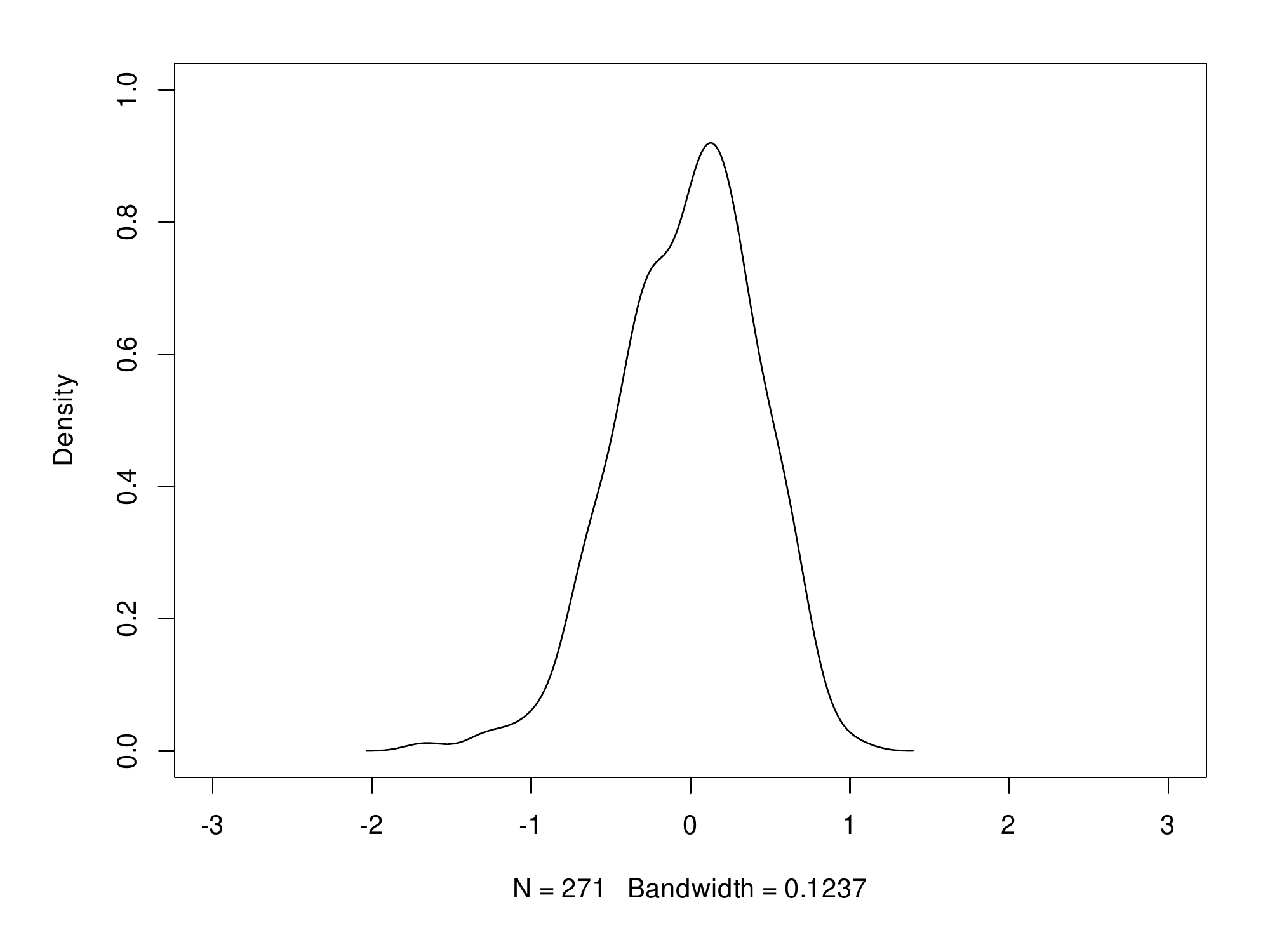}\\
 & & $p=0.01$ & $p=0.08$ &\\
 \hline
$D<1$\,Mpc &  $\log \rho_0 a_0^{3/2} \sim \mathcal{N}\!\left(a + b \, \log\!D + c \, \alpha,\sigma\right)$ & $W=0.99$ & $A=0.27$ & \includegraphics[width=2.5cm]{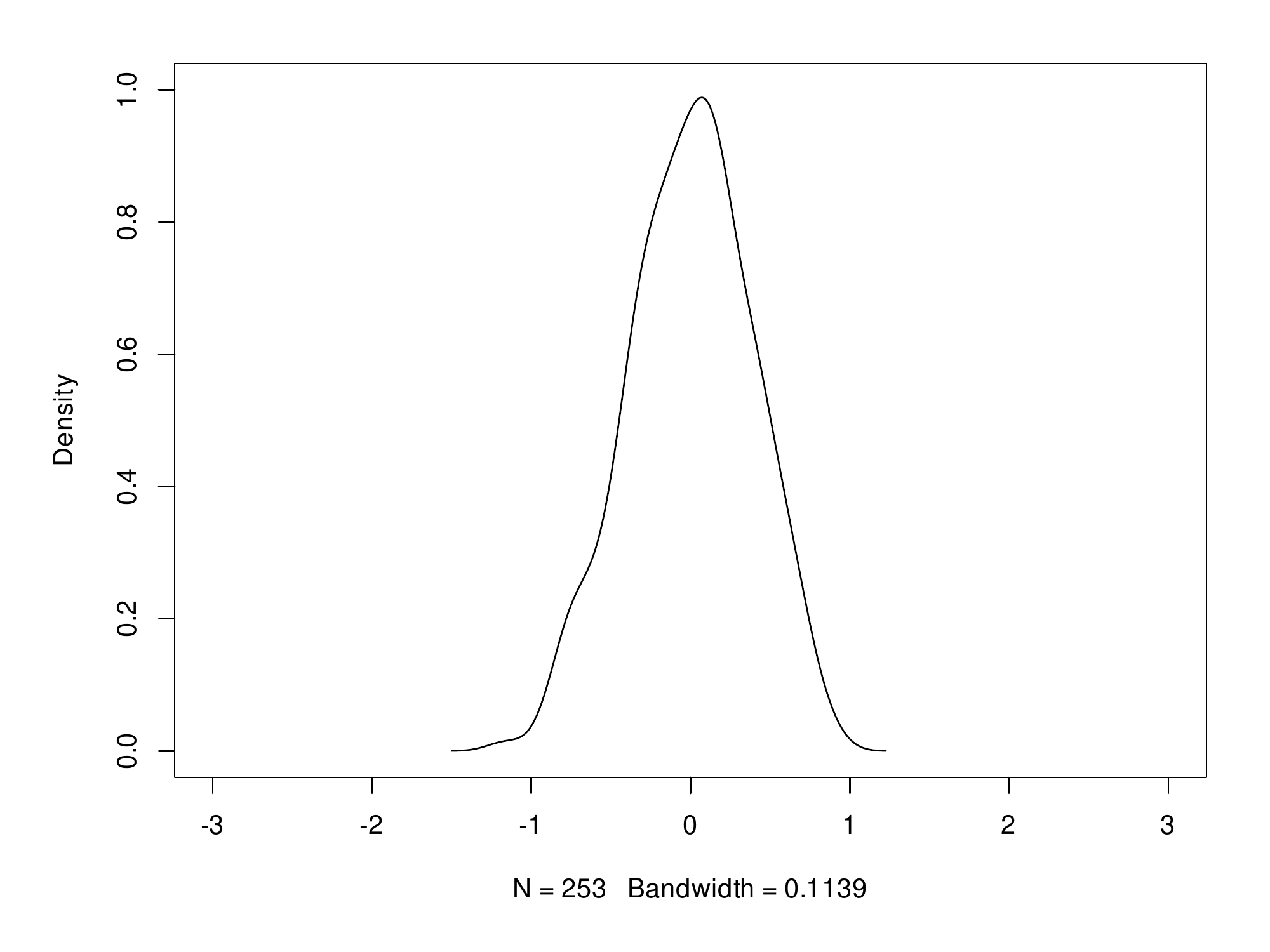}\\
 & & $p=0.38 $ & $p=0.68$ & \\
 \hline
full sample &   $\log \rho_0 a_0^{3/2} \sim \mathcal{N}\!\left(a + b \, \alpha;\,\sigma\right)$ & $W=0.99$ & $A=0.75$ & \includegraphics[width=2.5cm]{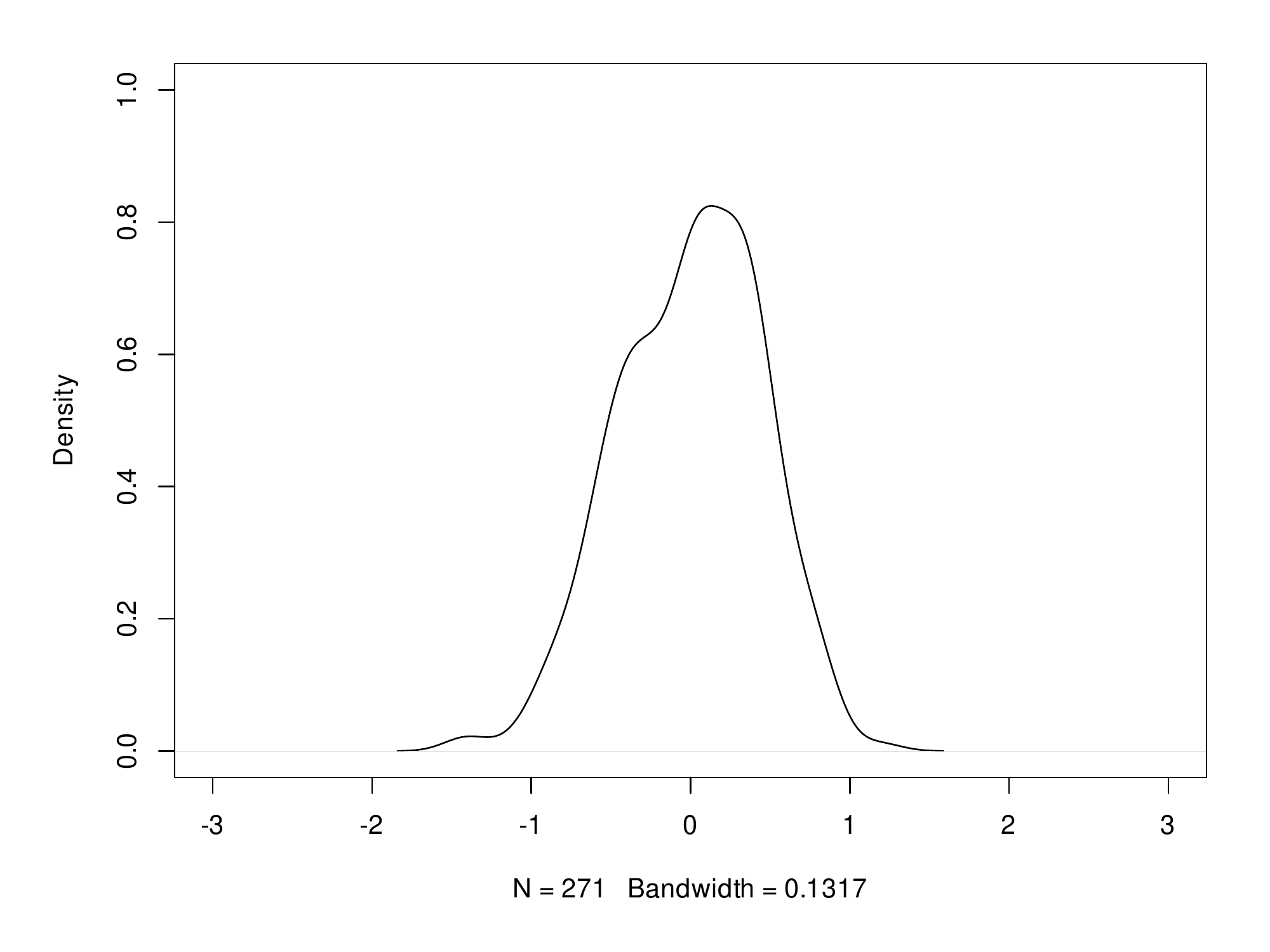}\\
 & & $p=0.10$ & $p=0.05$ & \\
 \hline
$D<1$\,Mpc & $\log \rho_0 a_0^{3/2} \sim \mathcal{N}\!\left(a + b \, \alpha;\,\sigma\right)$ & $W=0.99$ & $A=0.75$ & \includegraphics[width=2.5cm]{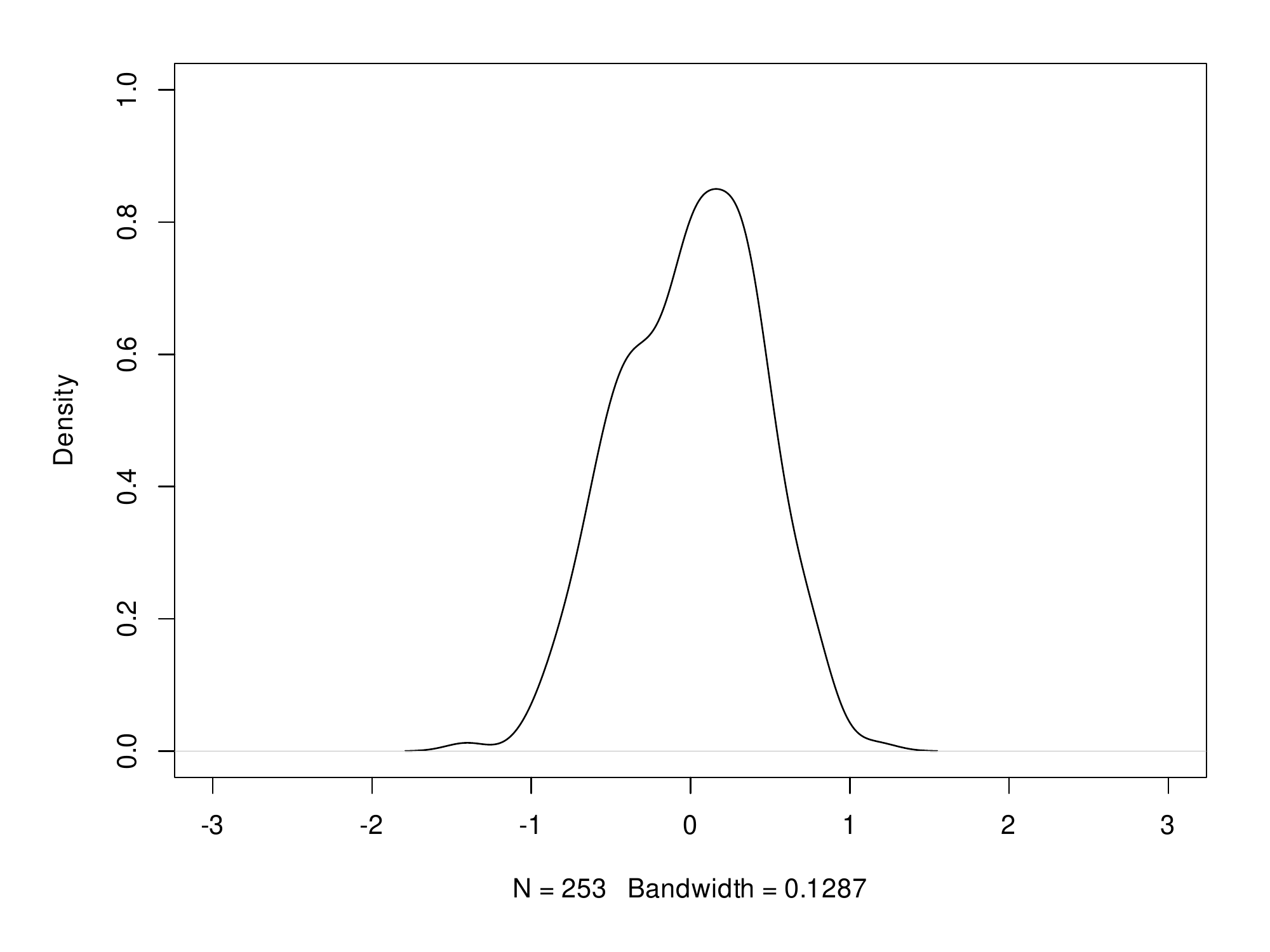}\\
 & & $p=0.15 $ & $p=0.05$ & \\
\hline
\enddata
\tablecomments{Col.(1) -- sample analyzed: ``full sample'' refers to 271 sources listed in Appendix\,\ref{ab}, while ``$D<1$\,Mpc'' refers to the sub-sample of 253 sources with linear sizes $D< 1$\,Mpc; Col.(2) -- statistical model considered; Col.(3) -- the value of the Shapiro-Wilk statistic and the corresponding p-value; Col.(4) -- the value of the Anderson-Darling statistic with the corresponding p-value; Col.(5) -- the residual density distributions resulting from fitting the regression line to the given dataset, x-axis range was set to (-3,3), and y-axis (0,1) for all plots.}
\end{deluxetable}

\end{document}